\documentclass[12pt,fleqn]{article}

\usepackage{amsmath, amssymb,graphicx,amsbsy}
\usepackage{color}
\usepackage{epsfig}
\usepackage{verbatim}
\usepackage{stmaryrd}
\usepackage{amsfonts}
\usepackage{young}
\input epsf

\newcommand \be{\begin{equation}}
\newcommand \ee{\end{equation}}
\newcommand \bea{\begin{eqnarray}}
\newcommand \eea{\end{eqnarray}}
\newcommand \bee{\begin{equation}}
\newcommand \eee{\end{equation}}

\newcommand\TL{\hfil$\displaystyle{##}$}
\newcommand\TR{$\displaystyle{{}##}$\hfil}
\newcommand\TC{\hfil$\displaystyle{##}$\hfil}

\def\seqalign#1#2{\vcenter{\openup1\jot
  \halign{\strut #1\cr #2 \cr}}}
\def\lbldef#1#2{\expandafter\gdef\csname #1\endcsname {#2}}
\newcommand{\eqn}[3][]{\lbldef{#2}{(\ref{#2})}%
\def\@eqnstyle{#1}%
\ifx\@eqnstyle\@empty%
\begin{equation} \eqalign{#3} \label{#2} \end{equation}%
\else%
\begin{equation} \seqalign{\span\TC}{#3} \label{#2} \end{equation}%
\fi}
\def\eqalign#1{\vcenter{\openup1\jot
    \halign{\strut\span\TL & \span\TR\cr #1 \cr
   }}}

\def\p2{{p \over 2}}
\def\half{{1\over 2}}
\def\nref#1{(\ref{#1})}

\input epsf

\begin{document}


\thispagestyle{empty}
\renewcommand{\thefootnote}{\fnsymbol{footnote}}

{\hfill \parbox{4cm}{
 ~~~
\\
  }}

\bigskip\bigskip

\begin{center} \noindent \Large \bf
Reflecting magnons
\end{center}

\bigskip\bigskip\bigskip

\centerline{ \normalsize \bf Diego  M. Hofman$^a$ and Juan M.
Maldacena$^{b}$\footnote[1]{\noindent \tt dhofman@princeton.edu ,
malda@ias.edu} }

\bigskip
\bigskip\bigskip

\centerline{$^a$ \it Joseph Henry Laboratories, Princeton
University, Princeton, NJ 08544, USA}
\bigskip
\centerline{$^b$ \it Institute for Advanced Study, Princeton NJ
08540, USA.}
\bigskip\bigskip

\bigskip\bigskip

\renewcommand{\thefootnote}{\arabic{footnote}}

\centerline{\bf \small Abstract}
\medskip
We study the worldsheet reflection matrix of a  string attached to a
D-brane  in   $AdS_5 \times S^5$. The D-brane corresponds to a
maximal giant graviton and it wraps an $S^3$ inside $S^5$. In the
gauge theory, the open string is described by a spin chain with
boundaries. We study an open string with a large $SO(6)$ charge,
which allows us to focus on one boundary at a time and to define an
asymptotic boundary reflection matrix. We consider two cases
corresponding to two possible relative orientations for the charges
of the giant graviton and the open string. Using the symmetries of
the problem we compute the boundary reflection matrix up to a phase.
These matrices obey the boundary Yang Baxter equation. A crossing
equation is derived for the overall phase. We perform weak coupling
computations up to two loops and obtain results that are consistent
with integrability. Finally, we determine the phase factor at strong
coupling using classical solutions.
  {\small
}

\newpage

\section{Introduction}
\label{INTRODUCTION}

Recently there has been a great deal of  progress in understanding
planar ${\cal N}=4$ super Yang Mills, see
\cite{Beisert:2005tm,Janik:2006dc,Beisertphase,
Arutyunov:2006iu,Hernandez:2006tk,Beisert:2005fw,afs}
and references therein. Planar Yang Mills theories give rise to a
two dimensional theory which can be viewed as the worldsheet of a
string. From the  gauge theory point of view, single trace operators
give rise to a closed spin chain, which in turn is related to a two
dimensional field theory on a circle. When the charges of the state
under consideration are very large one can view the gauge fixed
closed string theory \cite{fzam} as living on a large circle. The
limit where the string is infinite is particularly simple
\cite{Berenstein:2002jq,Staudacher:2004tk} and one can solve exactly
this problem
\cite{Beisert:2005tm,Janik:2006dc,Beisertphase,Santambrogio:2002sb}.
By ``solving''  we mean finding the fundamental excitations, their
dispersion relation,
 and
their scattering amplitudes on the infinite string for all values of
the 't Hooft coupling.
 It is very useful to consider the symmetries
of the problem, which are larger than naively expected
\cite{Beisert:2005tm}.
 These
symmetries determine completely the matrix structure of the two
particle scattering matrix \cite{Beisert:2005tm,Beisert:2006qh}.
 The remaining phase can then be determined by
using a crossing symmetry equation
\cite{Janik:2006dc,Beisertphase}.

In integrable field theories   it is often possible to define the
system on a half line, with suitable boundary conditions such that
the system remains integrable. A nice example is the boundary
Sine-Gordon theory studied in \cite{Ghoshal:1993tm}.
 In this article we study some physical
problems in ${\cal N}=4$ super Yang Mills that lead to a system
with a boundary. From the string theory point of view we expect to
have boundaries when we have D-branes. Then the open string
excitations are described by a two dimensional field theory with a
boundary. Such D-branes can arise in several situations:

\begin{itemize}
  \item  Gauge theories with additional flavors. Open strings
correspond to strings with a quark and an anti-quark at the ends.
  \item Theories with lower dimensional defects, which in some cases can
be realized as D-branes in the bulk \cite{DeWolfe:2001pq}.
  \item Certain large charge operators in ${\cal N}=4$ super Yang mills. For
example, operators of charge $N$ of the form $\textrm{det} (Z)$,
where $Z$ is one of the complex scalar fields in the theory. We will
focus on such operators and their excitations in this paper
\cite{McGreevy:2000cw,Balasubramanian:2001nh}.
\end{itemize}

Another  case where integrable systems with boundaries arise is when
we consider operator insertions along a Wilson loop
\cite{Drukker:2006xg}.   This is a situation where, despite the
absence of explicit D-branes in the bulk, we end up with a system
with a boundary. Of course, we could say that a Wilson line is a an
open string which ends on the boundary of $AdS_5  $.

Previous work analyzing open  spin chains  in ${\cal N}=4$ super
Yang Mills or the corresponding open strings with various boundary
conditions includes
\cite{DeWolfe:2001pq,Drukker:2006xg,Stefanski:2003qr,%
Chen:2004mu,Susaki:2004tg,mclias,DeWolfe:2004zt,
Berenstein:2005vf,Berenstein:2005fa,Erler:2005nr,
Okamura:2005cj,Susaki:2005qn,
Mann:2006rh,Berenstein:2006qk,Okamura:2006zr,Agarwal:2006gc,
Correa:2006yu,
de Mello Koch:2007uv}.   We focus, mainly, on
two intimately related cases which consist of giant graviton
operators with two possible orientations relative to the open
string ground state. We show that in one case we have boundary
degrees of freedom, while in the other case we do not.

The central idea in this paper is a  generalization of the
 analysis by Beisert \cite{Beisert:2005tm,Beisert:2006qh} to the case where we have boundaries.
Namely, we will use the symmetries of the system to determine the
matrix structure of the boundary scattering matrix. We then
proceed to write a crossing equation for the phase factor.
Although we have not solved the crossing equation, we have
computed the phase factor at weak and strong coupling.

  We have also checked that the boundary Yang
Baxter equation is obeyed. This follows by an argument similar to
the one used in \cite{Beisert:2006qh}. Furthermore, we performed
calculations at two loops in the weak coupling expansion and
obtained results compatible with integrability. At strong coupling,
this system
  leads to a classically integrable boundary condition
 for the string sigma model \cite{Mann:2006rh}.

When studying the action of the symmetries, it has proven to be
useful to have in mind the physical picture for the extra central
charges suggested by the classical string theory analysis in
 \cite{Hofman:2006xt} (see also \cite{Beisert:2006qh,Berenstein:2005jq} for a related picture).
  Although we explicitly discuss
the specific case of giant gravitons, our methods can be extended
without too much work to the various cases listed above.

This article is organized as follows. In section two we discuss
the boundaries related to giant gravitons, both in   string theory
and in the gauge theory.   In section three we derive
 the exact reflection matrices  up to an overall phase.
In section four we study these theories perturbatively in the weak
coupling limit. We   obtain the form of the phase factors up to
two loops and we also perform some explicit  checks of the exact
results. In section five we carry out an analogous discussion of
the strong coupling regime.
 We conclude with a discussion of these results in section six.
Finally, we include two appendices. In appendix \ref{apint} we
discuss the two loop integrability of the system with a boundary,
 while in appendix \ref{apr2} we present explicit
calculations of wave functions and reflection matrices at weak
coupling.

\section{Giant gravitons, determinants and boundaries}

We study open strings attached to maximal giant gravitons
\cite{McGreevy:2000cw} in
 $AdS_5 \times S^5$.
 These were previously studied at weak
coupling at one loop in \cite{Berenstein:2005vf}  and at two loops in
\cite{Agarwal:2006gc}, while a strong coupling
classical analysis was carried out recently in \cite{Mann:2006rh}.
Problems with the integrability of the theory at two loops were
pointed out in  \cite{Agarwal:2006gc}. We will see, however,
that a non trivial extra term coming from a subtle interaction
with the
boundary will render the theory integrable.\\

\subsection{Giant magnons meet giant gravitons}

\subsubsection{Giant gravitons}

Giant gravitons  are D3 branes in $AdS_5 \times S^5$
\cite{McGreevy:2000cw}. These D3 branes wrap topologically trivial
cycles, but are prevented from collapsing by their coupling to the
background fields. We will concentrate on the so called ``maximal
giant gravitons'' which are D3 branes wrapping a maximum size $S^3$
inside $S^5$. We can introduce coordinates for the $S^5$ in terms of
$W= \Phi^1 + i \Phi^2$, $Y= \Phi^3 + i \Phi^4$ and $Z= \Phi^5 + i
\Phi^6$, with $|Z|^2 + |W|^2 + |Y|^2 =1$. Maximal giant gravitons
are given by a pair of independent linear equations $a^I \Phi^I =b^I
\Phi^I =0$, and are all equivalent up to an $SO(6)$ rotation of the
sphere. These configurations preserve half of the supercharges. The
particular half that they preserve depends on their orientation
inside the $S^5$.

 We
are interested in studying open string excitations on the giant
gravitons. Our methods work best when the open string carries a
large amount of charge. Thus, we also want to single out a special
generator, $J = J_{56}$,  of $SO(6)$ which generates rotations in
the 56 plane. We consider open strings with large charge $J$. In the
field theory such states will involve a large number of insertions
of the field $Z$. Since we are breaking the SO(6) symmetry by
selecting a particular generator, $J$, we find that the explicit
open string description depends on the orientation of the giant
graviton inside $S^5$.

We will consider two cases where the D3 brane wraps the following
three spheres

\begin{itemize}

\item The three sphere given by $Z=0$. We will call this the
$Z=0$ giant graviton brane. We choose its orientation so that it preserves the same
supersymmetries as the field $Z$ in the field theory.

\item The three sphere given by $Y=0$, which we call the
$Y=0$ giant graviton brane. This brane preserves  half of the supersymmetries
preserved by the field $Z$ in the field theory.

\end{itemize}

\subsubsection{Giant magnons hitting giant gravitons}

\begin{figure}
\begin{center}
  \includegraphics[]{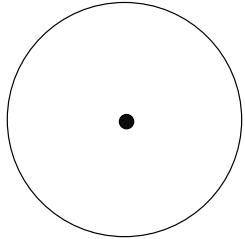}\\
  \caption{$Z=0$ brane in the $Z$ plane.}\label{gg1}
  \end{center}
\end{figure}

\begin{figure}
\begin{center}
  \includegraphics[]{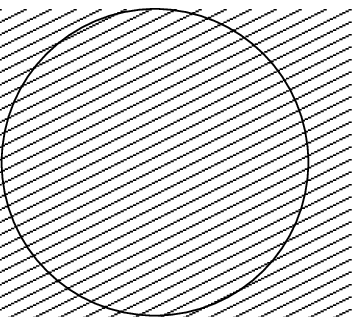}\\
  \caption{$Y=0$ brane in the  $Z$ plane.}\label{gg2}
  \end{center}
\end{figure}

 In what follows we will
study open strings with a large amount of charge $J$. The
centrifugal force   pushes most of
  this string to
 the circle at $|Z|=1$. We choose a light cone gauge so that
 a pointlike  string moving along this great circle
 corresponds to the BMN vacuum \cite{Berenstein:2002jq}.
 In light-cone gauge the string has length
 $J$.
 The ground state of this string preserves half of the
spacetime supersymmetries. In particular, it preserves those
supercharges with $\Delta -J=0$, where $\Delta $ is the conformal
dimension. Furthermore, we can have excitations with momentum $p$
 that move along the string. The lowest energy excitation with a given momentum
 is BPS. It corresponds to an elementary magnon on the corresponding gauge theory
 spin chain. The state manages to be BPS due to the existence of additional central
 charges \cite{Beisert:2005tm}.
  A convenient picture for the origin of these central
  charges is the following  \cite{Hofman:2006xt}.
 We draw the projection of the configuration on the $Z$ plane. This plane is
 embedded in $AdS_5 \times S^5$ as explained in detail in \cite{Lin:2004nb}.
 The string ground state corresponds to a point on the rim of the circle. An elementary
 excitation corresponds to a segment that joins two points on the rim. The two central charges
 correspond to string winding charges along this $Z$ plane \cite{Hofman:2006xt}.
 It is now convenient to think about the two branes mentioned above in these coordinates.
The $Z=0$ giant graviton brane is simply a point at $Z=0$, and it
wraps an $S^3$ inside the $S^5$, see figure \ref{gg1}. The $Y=0$
giant graviton brane, on the other hand, covers the whole disk, see
figure \ref{gg2}. At each point of the disk it also wraps an $S^1$
inside the $S^3$ that sits at that point. This circle shrinks at the
rim of the disk so that we end up with a brane with the  $S^3$
topology.

\begin{figure}
\begin{center}
  \includegraphics[]{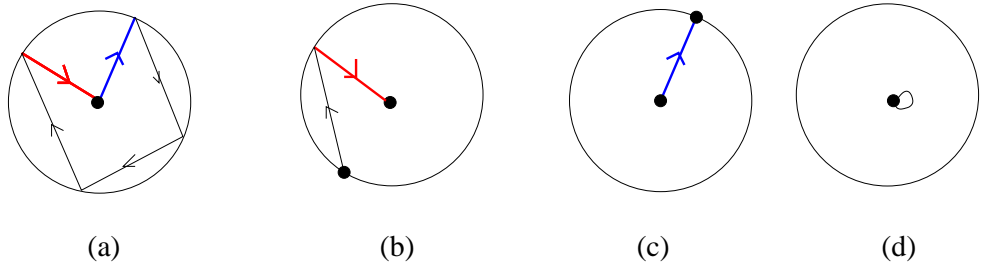}\\
  \caption{(a) Large $J$ open string attached to a $Z=0$ giant graviton brane.
  (b) Asymptotic form of the initial condition for the worldsheet scattering of a magnon off
  the right boundary. The dot on the boundary represents an infinite string in the lightcone ground
  state. (c) The boundary degree of freedom corresponds to a string going from the brane to
  the rim of the circle. (d) A string configuration for sufficiently small $J$ does not get close
  to the boundary of the circle. }
   \label{gg4}
  \end{center}
\end{figure}

In the large $J$ limit the string worldsheet is a very long
segment, so that when we analyze the effects near one of the
boundaries we can forget about the existence of the other boundary
and consider the system on a half infinite line. Therefore, we
consider first the problem of a giant magnon coming from infinity
and bouncing off the boundary back to infinity. In particular,
this means that our states interpolate between the usual vacuum of
BMN states \cite{Berenstein:2002jq} and the boundary. Furthermore,
this implies that one of the ends of the string looks like a
``heavy'' particle - i.e., there is an infinite amount of $J$
charge at this point - moving at the speed of light in a maximum
circle of $S^5$, see figure \ref{gg4} and \cite{Hofman:2006xt}.

Let us now look at the shape of the corresponding strings on the
$Z$ plane.
 The shape of this
string could be complicated at a random point in worldsheet time,
but in the asymptotic region (worldsheet time $t \rightarrow \pm
\infty$) they must look like giant magnons. This means they connect
two points on the rim of the  disk. This yields no surprise for the
$Y=0$ brane: the asymptotic scattering states for the $Y=0$ brane
are just strings stretched between points on the rim. This might
give the impression that the strings are contained within the
D-brane. This is not necessarily true; there is an additional $S^3
\subset S^5$ at each point on the disk and the brane and the string
could be separated within this $S^3$.

The $Z=0$ brane presents an interesting characteristic. In order for
the string to interpolate between the correct states we are led to
the following picture of the asymptotic scattering configuration,
see figure \ref{gg4} (b). We need to have a string that connects the rim
of the disk to the center where the $Z=0$ giant graviton brane sits.

This, in turn, suggests that the $Z=0$ brane carries a boundary
degree of freedom. Even  when there is not asymptotic excitation
we should have the piece of string connecting the rim of the disk
to $Z=0$, see figure \ref{gg4} (c).


A string lying along  a segment  in the $Z$ plane carries non
vanishing central charges of the worldsheet algebra, since we
argued that those central charges correspond to string winding
charges on
the $Z$ plane.

An important comment at this point is that strings with finite $J$
charge never reach the asymptotic vacuum described above and
consequently cannot reach the rim of the $Z$ plane. These strings
are localized around the brane at the center of the circle.

From the picture presented so far, we are lead to a simple guess for
the energy of the boundary state, once we understand the
representation of $SU(2|2)^2$ to which it belongs. Let us assume
that it belongs to the smallest BPS representation. We will later
substantiate this statement by a weak coupling computation where we
check that this is indeed the case. Once this is shown for weak
coupling, it will be true at all values of the coupling. This
implies that the energy is $\epsilon = \sqrt{ 1 + |k|^2}$ where $\vec k$
are the two
 the central charges.
We then notice that the central charge is precisely half the central
charge of a magnon with momentum $p=\pi$, which corresponds to a
string joining antipodal points on the rim. Therefore,
\begin{equation}
\epsilon_B = \sqrt{1 + 4 g^2} ~,~~~~~~~~~~~~~~g^2 = { \lambda \over 16
\pi^2 }
\end{equation}
\noindent where $\lambda$ is the 't Hooft coupling. Moreover, since
the string in figure \ref{gg4} (c)  is sitting at a point in the $S^3
\subset S^5$ we have collective coordinates and their quantization
is expected to lead to BPS boundary bound states with higher
$SU(2|2)^2$ charges, as we have
 in  the bulk \cite{Dorey:2006dq,Chen:2006gp}. These
states have energy $\epsilon_B(n) = \sqrt{n^2 + 4 g^2}$.

These statements do not rely on integrability, only on the
symmetries of the theory.
  Our exact and
perturbative calculations presented in the following sections
agree precisely with the results discussed above.

\subsection{Determinants in the gauge theory: the weak coupling
description}

The coordinates chosen in the previous section make it easy to
translate this analysis to the gauge theory side of the story. Here
we think of $W, Y, Z$ as the three complex scalars of ${\cal N}=4$
super Yang Mills (and of course we also have their complex
conjugates).

Then the $Z=0$ giant graviton brane, which is the maximal giant
graviton given by the equation $Z=0$, corresponds to the gauge
theory operator $\det(Z)$
\cite{Balasubramanian:2001nh,Corley:2001zk,Balasubramanian:2002sa,Berenstein:2004kk}.
 This is a gauge invariant operator with $J=N$.
Of course, the $Y=0$ giant graviton brane is then obtained by an
$SO(6)$ rotation as the operator $\det(Y)$. Both of these operators
correspond to the maximal giant gravitons on their ground state. We
now want to consider giant gravitons with open strings attached.
These are given by replacing one of the entries of the determinant
by a chain similar to the one appearing in single trace operators
\cite{Berenstein:2005vf,Berenstein:2005fa,Balasubramanian:2002sa,Berenstein:2002ke,
Berenstein:2003ah,Das:2000st,Balasubramanian:2004nb}.
 For example, for the $Y=0$  giant graviton brane we can
write
\begin{equation} \label{yoper}
\mathcal{O}_Y =\epsilon_{i_1 i_2 \ldots i_{N-1} B}^{j_1 j_2
\ldots j_{N-1}A}  Y^{i_1}_{j_1} Y^{i_2}_{j_2} \ldots Y^{i_{N-1}}_{j_{N-1}}
\left(Z Z Z \ldots Z Z Z\right)^B_A
\end{equation}

\noindent where one can make impurities propagate inside the chain
of $Z$s. Thus we consider operators of the form
\begin{equation}\label{oy}
\mathcal{O}_Y(\chi) = \epsilon_{i_1 i_2 \ldots i_{N-1} B}^{j_1 j_2
\ldots j_{N-1}A} Y^{i_1}_{j_1} Y^{i_2}_{j_2} \ldots
Y^{i_{N-1}}_{j_{N-1}} \left(\ldots Z Z Z \chi  Z Z Z
\ldots\right)^B_A
\end{equation}
where $\chi$ denotes a generic impurity. For the $Z=0$  giant graviton
brane, an
operator of the form (\ref{yoper}) with $Y$ replaced by $ Z$ would
factorize into a determinant and a single trace
\cite{Berenstein:2005fa}. This would not describe an open string but
a D-brane plus a closed string. Instead we consider excitations of
the form
\begin{equation}\label{oz}
\mathcal{O}_Z(\chi,\chi',\chi'') = \epsilon_{i_1 i_2 \ldots i_{N-1} B}^{j_1 j_2
\ldots j_{N-1}A}  Z^{i_1}_{j_1} Z^{i_2}_{j_2} \ldots
Z^{i_{N-1}}_{j_{N-1}}  \left( \chi Z Z \ldots Z Z Z \chi' Z Z Z
\ldots ZZ \chi''\right)^B_A
\end{equation}
where the impurities $\chi$ and $\chi''$ are stuck at  the ends of
the $Z$-string. The impurities will reflect when they get to the
ends of the string of $Z$s.  Of course, in the large $J$ limit, we
only have to worry about one of the ends at a time.


As we mentioned above the two kinds of giant gravitons are related
by an $SO(6)$ transformation. Thus, if we start with the $Z=0$ brane
and we add $Y$ impurities so as to completely ``fill'' the chain
 we would  end up with a state of the form
\begin{equation} \label{yzoper}
\mathcal{O}' = \epsilon_{i_1 i_2 \ldots i_{N-1} B}^{j_1 j_2
\ldots j_{N-1}A}  Z^{i_1}_{j_1} Z^{i_2}_{j_2} \ldots Z^{i_{N-1}}_{j_{N-1}}
\left(Y Y Y \ldots Y Y Y\right)^B_A
\end{equation}
which is simply an $SO(6)$ transform of the state $\mathcal{O}$ in
\nref{yoper}.

\section{Exact Results for the boundary reflection matrix}

Following the work of Beisert
\cite{Beisert:2005tm,Beisert:2006qh}, it is possible to calculate,
up to an overall phase,
the reflection matrix associated with the scattering of impurities
 from the boundaries discussed in the previous section.
  All we need are the symmetries of the theory and
the representations of the  states involved. In order to carry out
this analysis it is important to understand well the symmetries of
the system. Let us first discuss the symmetries of the bulk, before
we add the boundaries. As explained in
\cite{Beisert:2005tm,Beisert:2006qh} we have a centrally extended
$SU(2|2)^2$ algebra. We can consider one of these factors at a time.
Each factor has eight supercharges $\mathfrak{Q}^\alpha_{~a}$ and
$\mathfrak{S}^a_{~\alpha}$ which transform under $SU(2)\times SU(2)
\subset SU(2|2)$. We denote the generators of $SU(2)\times SU(2)$ as
$\mathfrak{R}^a_{~b}$,
$\mathcal{L}^\alpha_{ ~\beta}$ respectively.
We follow the notation of \cite{Beisert:2005tm}. The
algebra contains a generator $\mathfrak{C} ={ \epsilon \over 2} $,
where $\epsilon$ is the energy of an excitation around the vacuum
built with $Z$s, $\epsilon = \Delta - J_{56}$.
 In addition we have two extra
bosonic generators $k$ and $\bar k$ which are the extra central
charges which appear in the anti-commutators\footnote{ In the notation of
\cite{Beisert:2005tm} ${ k \over 2} = \mathfrak{P}$ and ${k^* \over 2} = \mathfrak{K}$.}
 \be
 \label{centralch}
 \{ \mathfrak{Q}^\alpha_{~a}, \mathfrak{Q}^\beta_{~b}\} =
 \epsilon_{ab}\epsilon^{\alpha \beta  } { k \over 2}
  ~,~~~~~~~~~~~~~
  \{ \mathfrak{S}^a_{~\alpha}, \mathfrak{S}^b_{~\beta}\} =
  \epsilon^{ab}\epsilon_{\alpha \beta  }
   { k^* \over 2 }
 \ee
These imply that the BPS condition reads $\epsilon^2 = 1 + k
 k^*$. For the fundamental bulk excitation  we also have a relation between
$k$ and the momentum \be  |k|^2 = 16 g^2 \sin^2 { p\over 2} \ee The
phase of $k$ is a bit more subtle and we will discuss it later.

\begin{figure}
\begin{center}
  \includegraphics[scale=1.4]{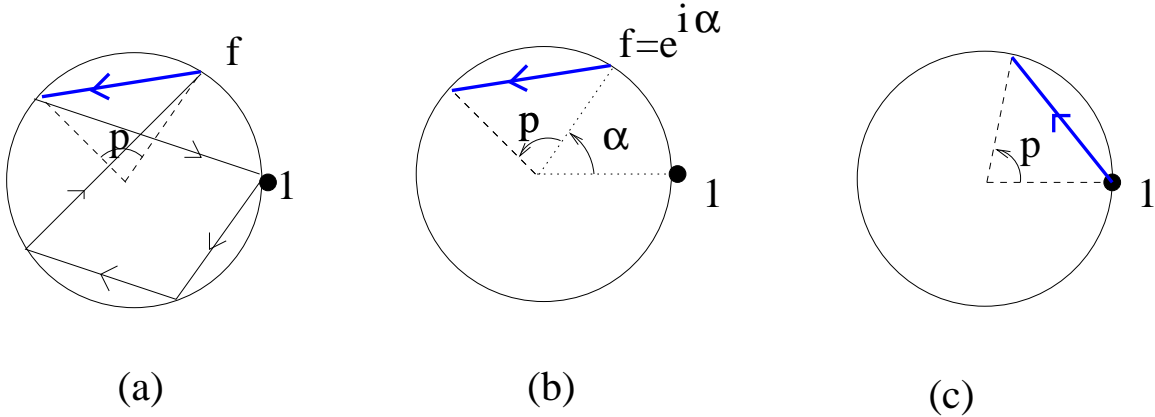}\\
  \caption{ (a) We depict a configuration of well separated magnons living on a long
  string. We choose the point 1 as a reference point and we want to describe the
  magnon with momentum $p$. (b) $f$ is the point on the unit circle where the
  magnon starts and gives the angle required to rotate it to the reference point 1, as in (c). }
  \label{fdefinition}
  \end{center}
\end{figure}

 The fundamental of
$SU(2|2)$ can be  split in the following way $\boxslash =
{}^B\boxempty \oplus \, {}^F\boxempty$, under $SU(2)\times SU(2)$,
 where we specified that one doublet is bosonic while
 the other one is fermionic, i.e.
${}^B\boxempty = (\phi^{\dot +},\phi^{\dot -})$ and
${}^F\boxempty = (\psi^{+},\psi^{ -})$. We have added a dot to the
bosonic $SU(2)$ indices to remind us that they transform under a
different $SU(2)$ than the fermions.
It is useful to write down the transformation rules for the fundamental multiplet as
 \begin{eqnarray}
 \mathfrak{Q}^\alpha_{~a} |\phi^b \rangle & =&  a \delta_a^b |
 \psi^\alpha \rangle ~,~~~~~~~~~~~~~~
 \mathfrak{Q}^\alpha_{~a} |\psi^\beta \rangle = b \epsilon^{\alpha \beta} \epsilon_{ab}
 | \phi^b \rangle
 \nonumber
 \\ \label{fundaction}
 \mathfrak{S}^a_{~\alpha} |\phi^b \rangle & =&  c
  \epsilon_{\alpha \beta} \epsilon^{ab} | \psi^\beta \rangle ~,~~~~~~~~~
 \mathfrak{S}^a_{~\alpha} |\psi^\beta \rangle = d \delta^\beta_\alpha
 | \phi^a \rangle
 \end{eqnarray}
where $ad - cb =1 $. We find that  ${ k \over 2 }  = ab$, $ { k^* \over 2 }  = cd$
and the energy is $\epsilon = 2 \, \mathfrak{C} =
ad + bc $.
We will pick the following parametrization for $(a,b,c,d)$:
 \begin{eqnarray} \notag
 a &=& \sqrt{g} \eta\\ \label{abcd}
 b &=& \frac{\sqrt{g}}{\eta} f \left(1-\frac{x^+}{x^-}\right)\\
 \notag
 c &=& \frac{\sqrt{g} i \eta}{f x^+}\\ \notag
 d &=& \frac{\sqrt{g}}{i \eta} \left(x^+ - x^-\right)
 \end{eqnarray}
 The momentum of the particle is
given by $\frac{x^+}{x^-} = e^{i p}$. The $ad - bc =1$ condition
translates into the mass shell condition
\begin{equation}
x^+ + \frac{1}{x^+} - x^- - \frac{1}{x^-} = \frac{i}{g}
\end{equation}
 \noindent The unitarity of the representation demands that
 \be \label{etadef}
 \eta = \sqrt{i x^- - i x^+}
 \ee
up to a phase, which we set to one.  Unitarity also requires that
 $f$
is  a phase, which contributes to the phase of the central charge as
$k = -2 g f ( e^{i p} - 1)$. We can think about the central charges
in terms of the segment that the magnon
describes in the $Z$ plane, by stretching from $z_1$ to $z_2$,
\be
  z_2-z_1=f( e^{i p} - 1) =  -{  k \over 2 g }
  \ee
  Then
the phase $f$ represents the orientation of that segment, see figure
\ref{fdefinition}.
 This orientation depends on the sum of the momenta of the magnons
that are to the left of the magnon under consideration\footnote{This corresponds to
the non-local parametrization of the problem, as described in
\cite{Beisert:2006qh}. This can also be described by forgetting about $f$ and
adding markers ${\cal Z}^\pm$, see \cite{Beisert:2006qh} for details. }.
 Thus $f$ is given by the angle that the magnon is making in a given
state, relative to the magnon with the same momentum that starts at $z_1=1$ and goes to
$z_2=e^{ip}$,
see figure \ref{fdefinition}. In the case that we have a semi-infinite string it is convenient
to take the reference point to coincide with the point where this infinite string is located on
the circle.

When we return to the full problem we need to consider two extended $SU(2|2)$
factors and the representation is the product of the fundamental
for each, giving a total of 16 states. For example we get
 \be \label{ywexpr}
 Y = \phi^{\dot -} \times  \tilde \phi^{\dot -}~,~~~~~~~
 W = \phi^{\dot +} \times  \tilde\phi^{\dot -}~,~~~~~~~~~{ \overline{ W}}
  = \phi^{\dot -} \times \tilde \phi^{\dot +}~,~~~~~~~~{ \overline{Y}}  =
 \phi^{\dot +} \times  \tilde \phi^{\dot +}
  \ee
   where the fields
$\phi^{\dot \pm}$ and $\tilde  \phi^{\dot \pm}$ transform under two
different $SU(2|2)$ groups. When we consider two extended $SU(2|2)$ factors
we get six central charges. However, in this physical problem
we require that the central charges for the two factors are equal (we set to zero the
difference).

When we consider the $Z=0$  giant graviton
brane we preserve the full symmetry
group. Physical states with finite $J$ correspond to strings that start and end
on the D-brane that sits at $Z=0$ and they thus carry zero total central charges $k=k^*=0$.

On the other hand when we consider the $Y=0$  giant graviton brane
we only preserve the subgroup which is also preserved by the field
$Y$. Let us consider the anticommutator
\be
\{ \mathfrak{Q}^\alpha_{~a},\mathfrak{S}^b_{~\beta} \} =
 \delta_a^b \delta^\alpha_\beta \mathfrak{C}
 + \delta^\alpha_\beta \mathfrak{R}^b_{~ a}
 + \delta_a^b \mathfrak{L}^\alpha_{~\beta}
\ee
 and concentrate on  the supercharges with a $\dot +$ index,
 $\mathfrak{Q}^\alpha \equiv
\mathfrak{Q}^\alpha_{~\dot +}$ and $\mathfrak{S}_{\alpha} \equiv
\mathfrak{S}^{\dot +}_{~\alpha} $. These supercharges annihilate an object with  $
\mathfrak{J} \equiv \mathfrak{C} +
\mathfrak{R}^{\dot +}_{~\dot +} = 0 $, which is a singlet under the second $SU(2)$,
such as a gauge invariant operator made purely with the field $Y$
(notice that an
upper $\dot -$ index carries $\mathfrak{R}^{\dot +}_{~\dot +}
=-\half $). These supercharges, together with $\mathfrak{J}$ and the second $SU(2)$
generators form an $SU(1|2)$ subgroup.
 The (noncompact) U(1) generator\footnote{
This factor is really non-compact in our problem, hopefully we can continue to call
it a $U(1)$ without causing confusion.}, $\mathfrak{J}$,  in
$SU(1|2)$, which appears in the right hand side of the supersymmetry algebra, is
given by
$ 2 \mathfrak{J} =     \epsilon    + 2 \mathfrak{R}^{\dot + }_{~\dot +} =
 \Delta - J_{56} - J_{34} - J_{12}  $ for one
$SU(1|2)$ factor and it is $2 \tilde {\mathfrak{J} } =   \epsilon
 +  2{\tilde{  \mathfrak{R}}}^{\dot +}_{~\dot +} =
  \Delta - J_{56} - J_{34} + J_{12}  $ for the other.

Let us now study each case in detail.\\

\subsection{The $Y=0$ giant graviton brane or  $SU(1|2)^2$ theory}
\label{exact1}

As we mentioned above, the symmetries that commute with the field
 $Y$ lead to an $SU(1|2)^2$ subgroup.
In order to study the problem we first focus on one $SU(1|2)$ subgroup and compute the reflection
matrix in this case.

The $SU(1|2)$ algebra arises by restricting all the generators of the $SU(2|2)$ algebra
to the ones carrying only ${\dot +}$ indices. As we mentioned above the (non-compact) U(1)
generator is $\mathfrak{J} = \mathfrak{C} + \mathfrak{R}^{\dot + }_{~ \dot +}$ and the
non-vanishing commutators are
\begin{eqnarray}
\left[\mathfrak{J}, \mathfrak{Q}^\alpha \right] &=& -\frac{1}{2} \mathfrak{Q}^\alpha\\
\left[\mathfrak{J}, \mathfrak{S}_\alpha \right] &=& \frac{1}{2} \mathfrak{S}_\alpha\\
\left[\mathcal{L}^\alpha_{~\beta}, \mathcal{J}^\gamma \right] &=&
\delta^\gamma_\beta \mathcal{J}^\alpha - \frac{1}{2}
\delta_\beta^\alpha \mathcal{J}^\gamma\\
\{\mathfrak{Q}^\alpha, \mathfrak{S}_\beta \} &=&
\mathcal{L}^\alpha_{~\beta} + \delta^\alpha_\beta \mathfrak{J}
\end{eqnarray}
\noindent where $\mathcal{J}^\alpha$ is any generator with upper
index $\alpha$. Notice that this algebra is not centrally extended.
All central extensions that appeared in the $SU(2|2)$ algebra  do
not contribute tot he anticommutators of the surviving supercharges have
disappeared. In this case a finite $J$ physical open string does not necessarily have zero
central charges, but the central charges, $k, ~k^*$ are not preserved by the boundary.

We can find the action of this algebra
on the states of the fundamental representation of $SU(2|2)$ from
\nref{fundaction}. For completeness we give the action of all generators
\begin{eqnarray} \nonumber
&\mathcal{L}^\alpha_{ ~\beta} |\phi^{\dot \pm} \rangle & = 0  ~,~~~~~~~~~~~~\quad \quad
\mathcal{L}^\alpha_{~\beta} |\psi^\gamma \rangle = \delta^\gamma_\beta
| \psi^\alpha \rangle - \frac{1}{2} \delta^\alpha_\beta
|\psi^\gamma\rangle\label{ac1}\\ \nonumber
&\mathfrak{J} |\phi^{\dot -}\rangle& = bc \, |\phi^{\dot -} \rangle
 ~,~~~~~\quad \quad \mathfrak{J} |\phi^{\dot +} \rangle
= ad \, |\phi^{\dot +} \rangle ~,~~~~\quad \quad \mathfrak{J} |\psi^\alpha \rangle =
\frac{1}{2}(ad+bc)\, |\psi^\alpha\rangle \label{ac2}\\ \nonumber
&\mathfrak{Q}^\alpha |\phi^{\dot -}\rangle & = 0~,~~~~~~~~~~~ \quad\quad \mathfrak{Q}^\alpha
 |\phi^{\dot +}\rangle  =
 a |\psi^\alpha\rangle ~,~~~~~~~ \quad\quad \mathfrak{Q}^\alpha |\psi^\beta\rangle  = b
\epsilon^{\alpha\beta} |\phi^{\dot -}\rangle \label{ac3}\\ \nonumber
&\mathfrak{S}_\alpha |\phi^{\dot -}\rangle & = c \epsilon_{\alpha\beta} |\psi^\beta\rangle
~,~~~~~~~~~\quad\quad \mathfrak{S}_\alpha |\phi^{\dot +}\rangle  = 0 ~,~~~~~~~~\quad\quad
\mathfrak{S}_\alpha |\psi^\beta\rangle  = d \delta^\beta_\alpha
|\phi^{\dot +}\rangle \label{ac4}
\end{eqnarray}
\noindent with $   \alpha,\beta,\gamma = +,-$.

Since the $SU(1|2)$ algebra does not have a central extension, we find that
for general momentum we   have a non-BPS representation since the charge
$\mathfrak{J} = { \epsilon \over 2} + \mathfrak{R}^{\dot +}_{~\dot +}$ can vary
continuously. Thus we expect that   the fundamental representation of
the extended $SU(2|2)$ transforms irreducibly.
In fact, it transforms as
 the representation of $SU(1|2)$ with the
supertableaux $\boxslash\!\boxslash $.
This has the right dimensions as
$\boxslash\!\boxslash =  {}^B1_{{ \epsilon \over 2} -{ 1 \over 2} }\,
\oplus {}^B1_{{ \epsilon \over 2}  + { 1 \over 2} } \, \oplus {}^F\boxempty_{\epsilon \over 2}$,
 where we have broken the
representation in $U(1) \times SU(2)$ multiplets and we have indicated whether we have bosons
or fermions.
In terms of the
  degrees of
 freedom of the $SU(1|2)$ fundamental representation $\boxslash = (\varphi, \chi^\pm)$ we
 can represent the corresponding states as $
 (\chi^+\chi^- -
\chi^-\chi^+, \varphi\varphi, \varphi\chi^\pm + \chi^\pm\varphi) $.
We now would like to match these states to the fundamental of the extended $SU(2|2)$ algebra.
 Matching their bosonic charges we see that
\begin{equation}
\boxslash\!\boxslash = \left(
                         \begin{array}{c}
                           \chi^+\chi^- -
\chi^-\chi^+ \\
                           \varphi\varphi \\
                           \varphi\chi^- + \chi^-\varphi  \\
                           \varphi\chi^+ + \chi^+\varphi  \\
                         \end{array}
                       \right) =\left(
                         \begin{array}{c}
                           \phi^{\dot -} \\
                           \phi^{\dot +}  \\
                           \psi^- \\
                           \psi^+ \\
                         \end{array}
                       \right)
                       \end{equation}


In the
 special case of  zero momentum $p=0$,   the representation
splits into two, one is the identity, given just by $\phi^{\dot -}$,
and the other three states form the fundamental, BPS representation
of $SU(1|2)$ with one bosonic, $\phi^{\dot +}$, and two fermionic
states. Recall that the field $Y $ is given by $Y= \phi^{\dot -}
\times \tilde \phi^{\dot -}$, so it is reasonable that for zero
momentum it is a singlet under $SU(1|2)$ since the $SU(1|2)$
subalgebra was found by demanding that all generators annihilate
$Y$. In this article  we are interested in the case with non-zero
momentum where we have a single $SU(1|2)$ non-BPS representation.

\subsubsection{The reflection matrix}

\begin{figure}
\begin{center}
  \includegraphics[scale=1]{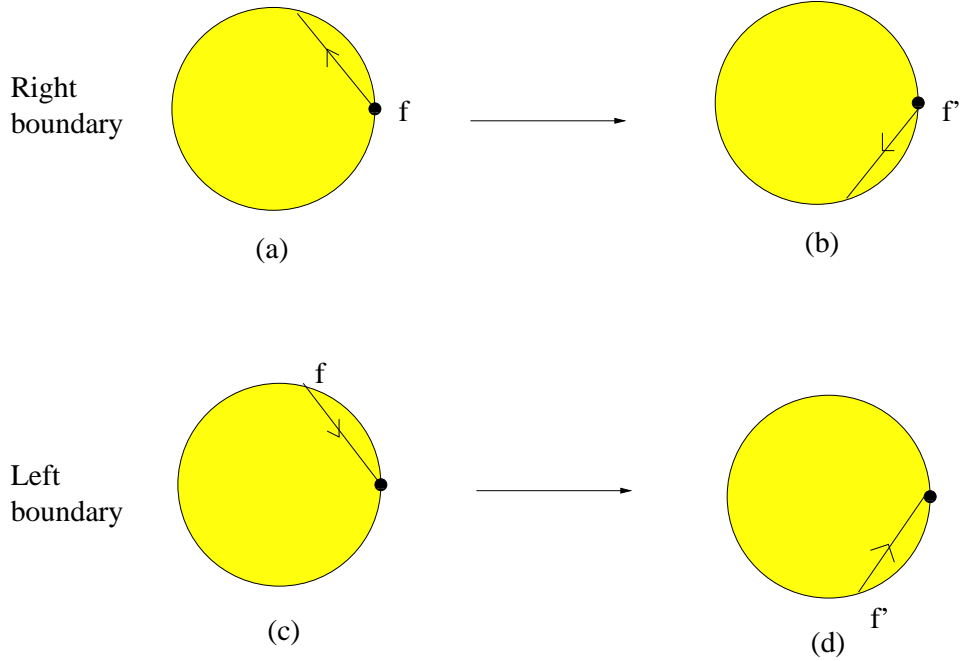}\\
  \caption{We depict several scattering configurations in a situation where we have
  a semi-infinite string. We choose the infinite region (``heavy'' particle / BMN vacuum) to lie at the reference point 1 in the complex plane.
  We can read off the values of the phase $f$ for the initial and final states from these
  figures. In (a) and (b) we depict the initial and final configuration for the scattering
  off a boundary on the right. We can see that in this case $f=f'=1$. In (c) and (d) we have the initial and final configurations
  for scattering from a boundary on the left. $f = e^{-i p} \neq f' = e^{+i p}$ in this setup. In all cases we located the point that sets the
  phase for the incoming state, $f$, and for the final state, $f'$. The arrow goes from
  left to right on the string worldsheet.}\label{conservation}
  \end{center}
\end{figure}

The $SU(1|2)$ reflection matrix\footnote{The full reflection matrix
of the theory is just the product of two $SU(1|2)$ reflection
matrices.}  $\mathcal{R}$ can now be calculated by demanding that
$\left[ \mathcal{R} , \mathcal{J}\right]=0$ for all generators
$\mathcal{J}$. The vanishing of the commutators of $\mathcal{R}$ and
the bosonic operators imply that $\mathcal{R}$ must be diagonal with
equal entries for the fermionic components. Namely,
\begin{equation}
\mathcal{R} = \left(
                \begin{array}{cccc}
                  r^- & 0 & 0 & 0 \\
                  0 & r^+ & 0 & 0 \\
                  0 & 0 & r & 0 \\
                  0 & 0 & 0 & r \\
                \end{array}
              \right)
              \end{equation}
The commutators with the fermionic operators yield the following conditions:
 \be
\begin{array}{rcl}
a r - a' r^+ &=& 0\\
b r^- - b' r &=& 0\\
c r - c' r^- &=& 0\\
d r^+ - d' r &=& 0
\end{array}
 ~~~~~~~~\longrightarrow~~~~~~~~~~
\begin{array}{l}
r^- = \frac{c}{c'} r = \frac{b'}{b} r\\
 ~~ \\
r^+ = \frac{a}{a'} r = \frac{d'}{d} r
\end{array} \label{equob}
\ee
\noindent where the primed variables are the quantum numbers of
the state after the reflection. These are obtained from the
original ones by
 \be \label{parity}
 x^\pm \rightarrow{x'}^\pm =  -x^\mp
 \ee
 This follows from conservation
of energy, $p \rightarrow -p$ and holding $x^+ + \frac{1}{x^+} -
x^- - \frac{1}{x^-} = \frac{i}{g}$. Note that $\eta$,
\nref{etadef}, is invariant under \nref{parity}, so $\eta'=\eta$. The phase $f$ might
change as well. $f$ represents the point where the magnon starts
in the $Z$ circle, see figure \ref{fdefinition}. When we have a
boundary scattering process the values for $f$ for the incoming
and the outgoing magnon are related by the geometry of the
scattering process in the $Z$ plane. In other words, it is
determined by the conservation laws. We represent the relevant
conservation laws in figure \ref{conservation} for the scattering
from a right boundary and a left boundary.

We see that in the case that we scatter from a boundary on the right, then
  $f$ does not change,   $f'=f$. If the orientation is opposite (boundary on the left), $f$
changes to $f' =  f \left(\frac{x^+}{x^-}\right)^2$, see figure \ref{conservation}(c,d).
Incidentally,  \nref{equob} requires $b c = b' c'$ and $a d = a' d'$. This
follows trivially from conservation of  energy $\epsilon = a d + bc$
and the mass shell condition $a d - b c =1$. Plugging in the values
for the quantum numbers yields

\begin{equation} \label{rightbdy}
\mathcal{R}_R = \mathcal{R}_{0 R}(p) \left(
                \begin{array}{cccc}
                  -e^{-i p} & 0 & 0 & 0 \\
                  0 & 1 & 0 & 0 \\
                  0 & 0 & 1 & 0 \\
                  0 & 0 & 0 & 1 \\
                \end{array}
              \right) ~,~~~~~~~~~{\rm for ~a ~right ~boundary}
              \end{equation}
\noindent  and
\begin{equation}\label{rl}
\mathcal{R}_L = {     \mathcal{R}_{0 L}(p) } \left(
                \begin{array}{cccc}
                  -e^{i p} & 0 & 0 & 0 \\
                  0 & 1 & 0 & 0 \\
                  0 & 0 & 1 & 0 \\
                  0 & 0 & 0 & 1 \\
                \end{array}
              \right)  ~,~~~~~~~~~{\rm for ~a ~left ~ boundary.}
              \end{equation}
In these   expressions
$\frac{x^+}{x^-} = e^{i p}$ and $\mathcal{R}_{0R}(p), ~ \mathcal{R}_{0L}(p)$ are  arbitrary
phases.
 We see that the two results are
consistent with the reflection symmetry that we have in the problem.
In fact, if we assume reflection symmetry we can also relate $\mathcal{R}_{0L}(p) =
 \mathcal{R}_{0R}(-p)$.
In addition, unitarity requires $\mathcal{R}_{0L}(-p) =1/\mathcal{R}_{0L}(p)$,
 $\mathcal{R}_{0R}(-p) =1/\mathcal{R}_{0R}(p)$.

The magnons in the full theory are the product of two fundamental magnons of each
extended $SU(2|2)$ algebra. Similarly, they are the product of representations for
each $SU(1|2)$ subalgebra.

From this result we can predict a ratio of reflection amplitudes.
For example the ratio of the amplitudes of scattering a
$Y = \phi^{\dot -} \times \tilde \phi^{\dot -} $ and a $W =
\phi^{\dot + } \times \tilde \phi^{\dot -}$ is $-e^{\mp i p}$ for $R , L$ boundaries respectively.
 Remember
that in our conventions $p$ is the incoming momentum. If the boundary is placed on the left
this momentum is negative. So left and right results are consistent.
 We will compare this result with explicit calculations in the following sections.

Another interesting comment is that this matrix does not contain
poles or zeros, unless they are included explicitly in
$\mathcal{R}_0(p)$. This means that if there is a bound state in one
channel, all channels must have one. In the next section we will
check that there is no bound state at weak coupling. We will also
compute  $\mathcal{R}_0(p)$ perturbatively to two loops at weak
coupling and to leading order at strong coupling.

\subsubsection{The Yang Baxter equation}
\label{exact1YB}

\begin{figure}
\begin{center}
  \includegraphics[scale=0.4]{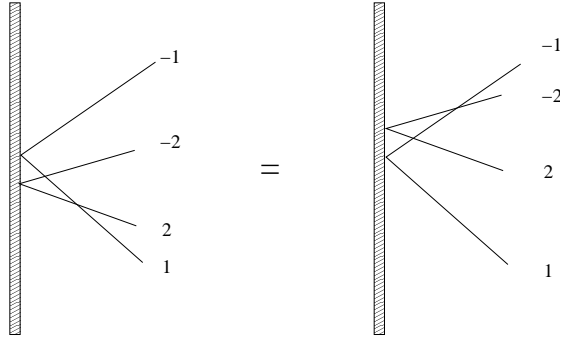}\\
  \caption{The content of the
   Yang Baxter equation is that these two processes give the same answer.}\label{ybe3fig}
  \end{center}
\end{figure}

We now check that this
reflection matrix satisfies the boundary Yang Baxter equation. This
equation is represented graphically in figure \ref{ybe3fig} and it
states that one can compute the reflection of a pair of particles in
two ways. As in the case of the bulk Yang Baxter equation one can
check the equation in a simple way using the symmetries
\cite{Beisert:2006qh}. The idea is to look at the Hilbert space of
two particles and decompose it in representations of $SU(1|2)$ and
then check the equation in each representation. This can be done in
a simple way if each representation contains a state that scatters
diagonally, so that all scattering amplitudes are simply phases.
The intermediate representations of
the 2 particle incoming states are:
\begin{equation}\label{ybox}
 \begin{Young} $\diagup$ & $\diagup$  \cr
     \end{Young} \times \begin{Young}
$\diagup$ & $\diagup$  \cr
     \end{Young}  =  \begin{Young} $\diagup$ & $\diagup$  & $\diagup$ & $\diagup$ \cr
     \end{Young} +  \begin{Young} $\diagup$ & $\diagup$  & $\diagup$
     \cr $\diagup$ \cr
     \end{Young} + \begin{Young} $\diagup$ & $\diagup$
     \cr $\diagup$ & $\diagup$ \cr
     \end{Young}
     \end{equation}
The first representation on the right hand side of (\ref{ybox})
contains the state $\phi_1^{\dot +} \phi^{\dot +}_2$, the second contains the states
$\psi_1^+ \psi_2^+$ and $\psi_1^{-} \psi_2^{-}$
 and the third one contains $\phi_1^{\dot -}  \phi_2^{\dot -}$, which are all
 states that scatter diagonally.

Let us now check the boundary Yang Baxter equation for two excitations that scatter diagonally.
Let us denote by $S(1,2)$ their bulk scattering. $S(1,2)$ is simply a phase by assumption.
Similarly, we have the reflection $r(1)$ and $r(2)$ from the boundary which is also a phase.
Thus we have
\begin{equation}
S(1,2)   r(1)  S (-2,1) r(2) =  r(2) S(-1,2) r(1) S(-2,-1)
\end{equation}
Since we only have phases we see that $r(1)$ and $r(2)$ drop out from the equation and we are
only left with a  requirement involving the bulk $S$ matrix. This requirement is obeyed if
the bulk $S$ matrix is   parity invariant,
$S(1,2) = S(-2,-1)$. This is an invariance of the bulk S matrix, thus we see that
the boundary Yang Baxter equation is satisfied. We have also checked explicitly that the
equation is indeed satisfied.

\subsubsection{The crossing equation}\label{crosssect}

In order to derive the crossing equation we need to form a singlet
state according to the derivation in \cite{Beisert:2006qh}. This
identity state is

\begin{eqnarray} \label{identsta}
{\bf 1}_{(p, \bar p )} &=&
  f_p e^{i p/2}
 ( \phi_p^+ \phi_{\bar p }^- - \phi_p^- \phi_{\bar p }^-) + \epsilon_{\alpha \beta}
 \psi_{p}^\alpha \psi_{\bar p}^\beta
\end{eqnarray}
where the subindex $p$  denotes the momentum and energy
$\epsilon(p)$ of the first particle and the index $\bar p$ denotes
the momentum $ \bar p =  -p$ and energy $\bar \epsilon = -
\epsilon(p)$ of the second, crossed, particle.
 If we think in terms of the fermionic part of the
state we can view the state as a hole, $\psi_+(p)$, and
  negative energy
electron $\psi_-(\bar p)$. In this case, we clearly see that we get back the original vacuum
of the theory. Thus adding this state should have no effect on the theory.
By scattering this two particle state from a third and demanding that
the result is  invariant one can obtain the crossing equation
\cite{Janik:2006dc,Beisert:2006qh}.

\begin{figure}
\begin{center}
  \includegraphics[scale=0.6]{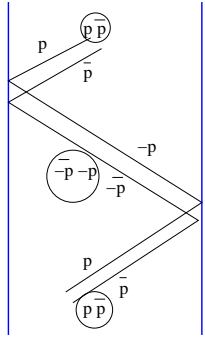}\\
  \caption{ We scattering the singlet state $p\bar p$ from the right boundary and then
  from the left boundary in order to come back to the original situation. We demand that
  this double scattering gives one.   } \label{crossing}
  \end{center}
\end{figure}

If  we start with this state and we scatter it from the right
boundary we  obtain the state $ r(p) {\bf 1}_{(-\bar p, - p)}$, where $r(p)$ is
some reflection phase. We
see that we do not get the same state because the particle and
antiparticle are in a different order. However, if we have a left
boundary and we now scatter the resulting state we get back to the
original state \nref{identsta}, see figure \ref{crossing}. We now use that parity invariance
implies that the scattering phase we get from the second scattering
 is the same as the one we got from the
first boundary. Thus we find that the total scattering phase is
$r(p)^2$. Now it makes sense to demand that the total scattering
phase is one, $r(p)^2 =1$.

 So, we get  $r(p) = \pm 1$. By considering different boundaries on the two sides
we see that the signs should be all plus or all minus, for all
boundaries in the theory. We take this sign to be plus. We'll show
this in a moment, by looking at the plane wave limit.

When we scatter this state from the boundary we will need the boundary reflection matrix
\nref{rightbdy} and the bulk S matrix written in \cite{Beisert:2006qh}.

At the end of the day we obtain
\begin{eqnarray}
{\bf 1}_{(p ,\bar p) } &=& h_b S_0(p,-\bar p) {\cal R}_{0R}(p) {\cal
R}_{0R}(\bar p) \left[ f_{-p} e^{i p/2}
 ( \phi_{ - \bar p}^+ \phi_{- p }^- - \phi_{- \bar p}^- \phi_{- p }^-) +
 \epsilon_{\alpha \beta}
 \psi^\alpha_{- \bar p}  \psi^\beta_{-p} \right] = \notag
 \\ \nonumber
 &=&  h_b  S_0(p,-\bar p) {\cal R}_{0R}(p) {\cal R}_{0R}(\bar p) { \bf 1 }_{(-\bar p , -p)}
\\
h_b &\equiv&  {{ 1 \over x^-}   + x^- \over { 1 \over x^+} + x^+ }
\end{eqnarray}
where   $S_0$ is the phase factor as defined by Beisert in
\cite{Beisert:2006qh} and ${\cal R}_0$ is the phase factor which
multiplies the boundary reflection matrix that we had above. Thus
the crossing equation has the form \be
 {\cal R}_{0R}(p) {\cal R}_{0R}(\bar p) = { 1 \over h_b} { 1 \over S_0(p, -\bar p)}
 =
 {{ 1 \over x^+} + x^+ \over {1 \over x^-}  + x^- } { 1 \over S_0(p, -\bar p)}
 \ee
This would be the equation in the case that we had only one $SU(2|2)$ factor.
In the full theory, where we have the two $SU(2|2)$ factors we
define the full reflection factor to be simply ${\cal R}_{0R}^2(p)$,
and the bulk phase factor is usually written in terms of a dressing
factor $\sigma^2 $ through the equation \cite{afs} \be S_0(p_1,p_2)^2 =
{ (x^+_1 - x_2^-) \over(x^-_1 - x_2^+)}{( 1 - { 1 \over x_1^- x_2^+
} ) \over ( 1 - { 1 \over x_1^+ x_2^- } ) } { 1 \over \sigma^2(p_1,p_2)
} \ee Then the equation for the full theory becomes

\begin{equation}\label{crossfin}
 {\cal R}^2_{0R}(p) {\cal R}^2_{0R}(\bar p) ={
1 \over h_b^2} { 1 \over S^2_0(p, -\bar p)} =
  { x^+  + { 1 \over x^+ } \over x^- + { 1 \over x^- }}
 {   \sigma^2(p,- \bar p) }
\end{equation}

Notice that in the plane wave limit \cite{Berenstein:2002jq}
the right hand side of this
equation is just 1. In this limit our theory is non interacting and
we know that, in the $SU(2)$ subsector, ${\cal R}^2_{R}(p) = {\cal
R}^2_{R}(\bar p)=-1$, as this is just a relativistic theory with
Dirichlet boundary conditions. From equation \nref{rightbdy} we see
that this implies ${\cal R}^2_{0R}(p) = {\cal R}^2_{0R}(\bar p)=-1$.
This means that the plus sign is the correct one for the right hand
side of equation \nref{crossfin}.

Finally, we should also mention that unitarity implies
 \be
 {\cal R}_{0R}(p) {\cal R}_{0R}(-p)   =1
 \ee

\subsection{The $Z=0$ giant graviton brane or  $SU(2|2)^2$ theory}

We now study the case of a $Z=0$ giant graviton brane, which
preserves the full $SU(2|2)$ symmetry,  see figures \ref{gg1},
\ref{gg4}. The new feature of this case is the existence of a
boundary degree of freedom. We assume that the boundary degree of
freedom transforms in the fundamental representation of extended
$SU(2|2)^2$. It seems clear that this is the case at weak coupling
where we have an impurity stuck between the $Z$-determinant and
the string of $Z$'s producing the large $J$ open string. Then we
expect that this should continue to be the case at all values of the
coupling. Since the supersymmetry algebra has been extended by the
addition of two central charges we need to understand the values of
the central charges for the impurity. Here, we will be guided by the
string pictures we discussed above, where the central charges are
associated to the winding number of the string in the $z$ plane.
Thus the central charge vector is simply the vector given by a
string going from the brane at $z=0$ to the rim of the disk, see
figure \ref{gg4} (c). We can also view the central charge vector as a
complex number. This fixes the absolute value of the central charge
vector \be |k|^2 = 4 g^2 \ee
 The phase of the central charge depends on the
momenta of  the other magnons that are in the problem and changes
when a magnon scatters from the boundary. Below we will  explain how
it changes. The conclusion is that the representation of the
boundary impurity is again the fundamental of the extended symmetry
algebra. The only difference between the impurity representation and
the magnon one is in the relation between the central charges and
the momentum (the impurity does not have a momentum quantum number),
and in the precise dynamics of the phase of the central charge. It
turns out that the problem completely factorizes into each extended
$SU(2|2)$ factor. Thus we
 consider first the case where we have only one $SU(2|2)$ factor.

\begin{figure}
\begin{center}
  \includegraphics[scale=0.8]{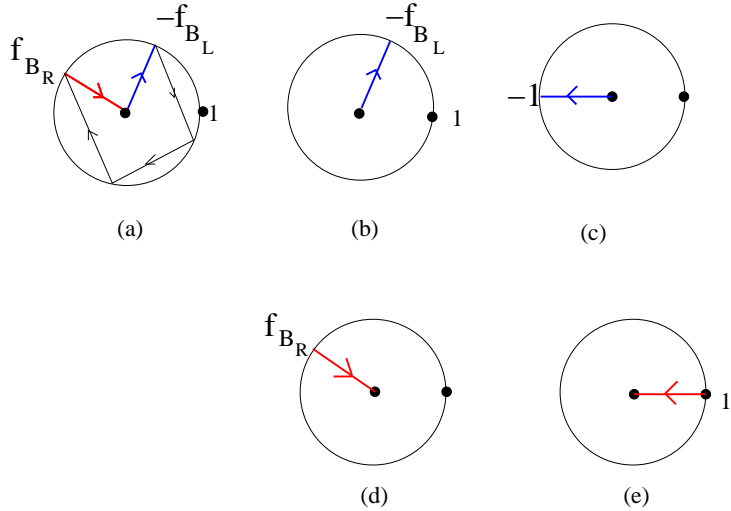}\\
  \caption{ In (a) we see a generic open string configuration in the regime that $J$ is very large and
  the magnons are very well separated. We have denoted by $f_{B_L}$ and $f_{B_R}$ the corresponding
  parameters of left and right boundaries, respectively.
   In (b) we isolate the piece of string corresponding to the
  left boundary impurity. Its phase $-f_B$ is the end point of this string. $f_B$ is also
  the phase by which the configuration was rotated with respect to
  the  reference configuration in (c).  In (d) we isolated the piece of string corresponding
  to the right boundary. $f_B$ is the starting point of the string on the circle. This phase
  is also
  the one by which the configuration was rotated with respect to
  the  reference configuration in (e).
  These figures can be viewed as the central charge vectors (except for a $-2g$ factor)
  for the states involved and also as the projections of the
  physical string configurations
   to the $z$ plane in the $AdS_5 \times S^5$ geometry.
  }\label{fbdydef}
  \end{center}
\end{figure}

Let us start by being more specific about the representation
properties of the boundary degree of freedom. The transformation
properties are as in the bulk case, \nref{fundaction}, but with the
following values of $a,b,c,d$.
\begin{eqnarray}
a_B &=& \sqrt{g} \eta_B\\
b_B &=& \frac{\sqrt{g} f_B}{\eta_B}\\
c_B &=& \frac{\sqrt{g} i \eta_B}{x_B f_B}\\
d_B &=& \frac{\sqrt{g} x_B}{i \eta_B}
\end{eqnarray}
where we have added the subindex $B$ to distinguish these from the
bulk case.  Unitarity of the representation requires $|\eta_B|^2 = -
i x_B$ and that $f_B$ is just a phase. The shortening/mass shell
condition implies
\begin{equation}
ad-bc =1 \quad \longrightarrow \quad x_B + \frac{1}{x_B} =
\frac{i}{g} ~,~~~~~~~x_B = \frac{i}{2g} \left(1+\sqrt{1+4g^2}\right)
\end{equation}
where we picked the solution for $x_B$ which leads to positive
energy
\begin{equation}\label{enB}
\epsilon = ad +bc = \frac{g}{i} \left(x_B - \frac{1}{x_B}\right) =
\sqrt{1 + 4g^2}
\end{equation}
The phase $f_B$ depends on the other magnons in the problem and can
be understood most simply by looking at figure \ref{fbdydef}. For a
right boundary, $f_B$ is the position of the endpoint of the last
magnon on the $Z$ circle. Equivalently it is given by the sum of the
momenta of all magnons to the left of the boundary. Since the system
ends at the right boundary, this means that $f_B= \prod_{j} e^{i
p_j} f_1$ for all the magnons in the system, where $f_1$ is the
starting point of the first magnon.

\begin{figure}
\begin{center}
  \includegraphics[scale=0.8]{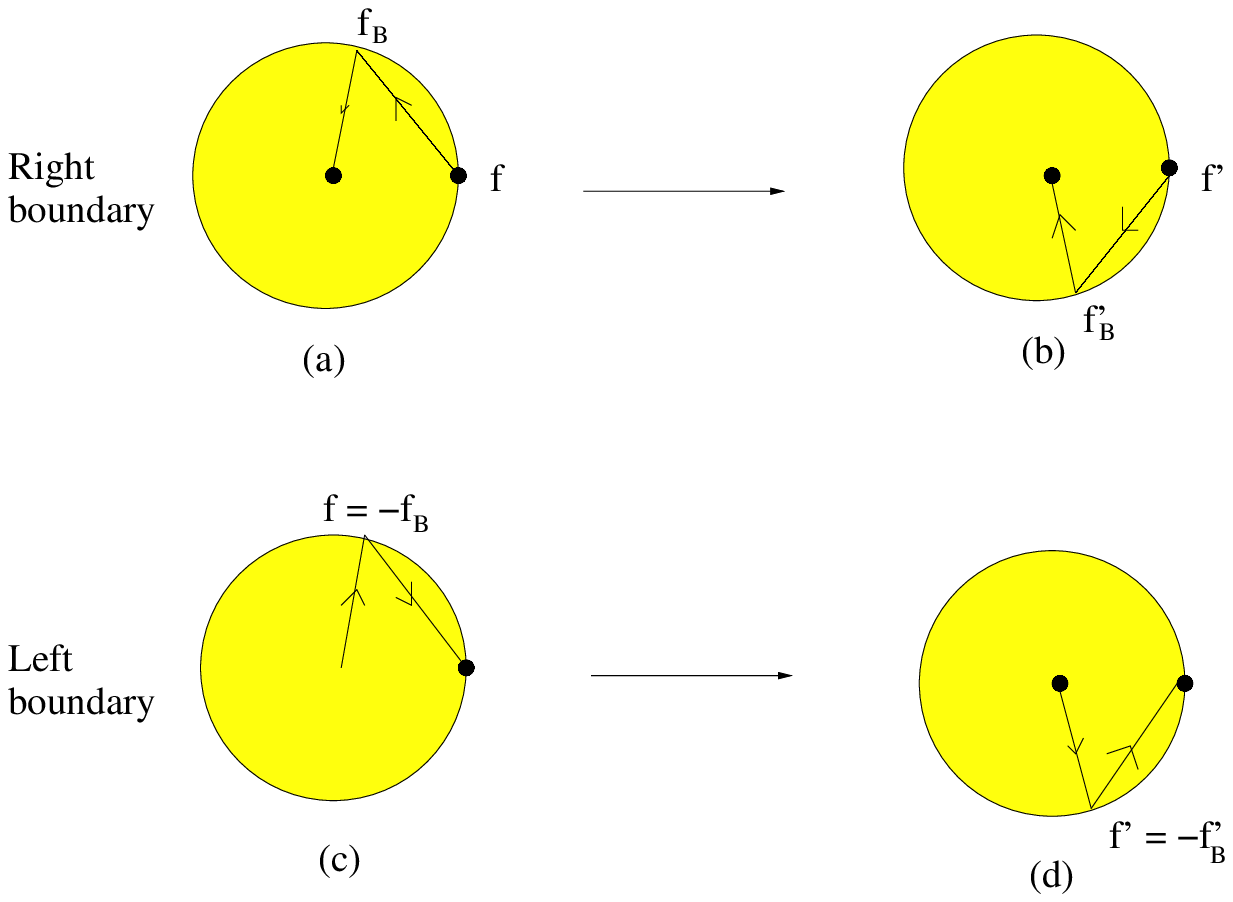}\\
  \caption{ In (a) we see the initial state for scattering from a right boundary and in (b)
  we see the final state. We have indicated the phases of the central charge in both cases.
  In (c) we see the initial state for right boundary scattering and in (d) we see the final state.
  These figures can be viewed as the central charge vectors (except for a $-2g$ factor) for
   the states involved and also as the projections of the
  physical string configurations  to the $z$ plane in the $AdS_5 \times S^5$ geometry.
  }\label{fbdyscattering}
  \end{center}
\end{figure}

We now derive the boundary $S$ matrix for this system. We must first
understand how $f$ and $f_B$ change under scattering, see figure
\ref{fbdyscattering}. Let us consider the case of right boundary
scattering.  In the initial state we have $f_B = e^{ip} f$. In the
final state the magnon phase does not change, $f'=f$ and $f'_B =
e^{-ip } f= e^{ - 2 i p} f_B = \left(\frac{x^-}{x^+}\right)^2 f_B$,
see figure \ref{fbdyscattering} a, b. On the other hand, for a left
boundary  $f_B = -f$, see figure \ref{fbdyscattering} c, d. In this
case $ f'=-f'_B = e^{2 i p }f $, or  $f'_B =
\left(\frac{x^+}{x^-}\right)^2 f_B$. $x_B$ does not change in either
case.

Let us now analyze the case with a left boundary in detail. The
following equations summarize the quantum numbers of the incoming
particle and the boundary and how they change after scattering:

\be
\begin{array}{rclcrcl}
a &=& \sqrt{g} \eta  & ~~~~ ~~~~~~~~ &   a &=&a\\
b &=& \frac{\sqrt{g}}{\eta} f \left(1-\frac{x^+}{x^-}\right) &
~~~~~~~~~~~~ &  b' &=&
 - \frac{x^+}{x^-} b\\
c &=& \frac{\sqrt{g} i \eta}{f x^+}   & ~~~~~~~~~~~~ &  c' &=& -\frac{x^-}{x^+} c\\
d &=& \frac{\sqrt{g}}{i \eta} \left(x^+ - x^-\right)  & ~~~~~~~~~~~~
& d'&=&d
\end{array}
\ee
\be
\begin{array}{rclcrcl}
a_B &=& \sqrt{g} \eta_B  & ~~~~ ~~~~~~~~ &
 a_B' &=&  a_B\\
b_B &=& -\frac{\sqrt{g} f}{\eta_B}  & ~~~~ ~~~~~~~~ &
 b_B' &=&  \left(\frac{x^+}{x^-}\right)^2 b_B\\
c_B &=& -\frac{\sqrt{g} i \eta_B}{x_B f}  & ~~~~ ~~~~~~~~ &
 c_B' &=&  \left(\frac{x^-}{x^+}\right)^2 c_B\\
d_B &=& \frac{\sqrt{g} x_B}{i \eta_B}  & ~~~~ ~~~~~~~~ &
 d_B' &=& d_B
\end{array}
\ee


In order to calculate the reflection matrix, $\mathcal{R}$, we
demand that all commutators of the reflection matrix with the
generators of $SU(2|2)$ vanish. In this case the operators act on
two particle states, so the computation is more involved that in the
last case. In particular, we have to remember that fermionic
operators acting on two particle states are defined as $Q = Q_1
\otimes 1 + (-)^F \otimes Q_2$, where $F$ is the fermionic number of
particle state 1. The computation is almost identical  to the one
performed in \cite{Beisert:2005tm}. Invariance under the bosonic
generators implies that the ${\cal R}$ matrix can be written as
 \cite{Beisert:2005tm}\cite{Beisert:2006qh}
\begin{eqnarray}
 \mathcal{R}
 |\phi_B^a \phi_p^b\rangle &=& A |\phi_{B}^{\{a} \phi_{-p}^{ b\} } \rangle +
 B |\phi_{B}^{[a}\phi^{ b]}_{-p} \rangle + \frac{1}{2}
 C \epsilon^{a b}\epsilon_{\alpha \beta}
 |\psi_{b}^\alpha \psi^\beta_{-p}
\rangle\\
\mathcal{R} |\psi_B^\alpha \psi_p^\beta \rangle &=& D
|\psi_{B}^{\{\alpha}\psi_{-p}^{ \beta\}} \rangle + E
|\psi_{B}^{[\alpha} \psi_{-p}^{ \beta]}\rangle + \frac{1}{2} F
\epsilon^{\alpha
\beta}\epsilon_{a b} |\phi_{B}^a \phi_{-p}^ b \rangle\\
\mathcal{R} |\phi_B^a \psi_p^\alpha \rangle &=& G |\psi_{B}^\alpha
\phi_{-p}^a \rangle +
 H |\phi_{B}^a \psi_{-p}^\alpha
\rangle\\
\mathcal{R} |\psi_B^\alpha \phi_p^a \rangle &=& K |\psi_{B}^\alpha
\phi_{-p}^a \rangle + L |\phi_{B}^a \psi_{-p}^\alpha \rangle
\end{eqnarray}
\noindent where $a,b$ represent bosonic indices, $\dot \pm$, and
$\alpha, \beta$ are fermionic indices, $\pm$. The (anti)
symmetrization symbols are defined with a $\frac{1}{2}$
normalization factor, i.e. $\{a b\} = \frac{ab + ba}{2}$.

It is understood that the states on the right hand side of these
equations are out states and, therefore, have primed quantum
numbers. In particular, they have primed phases, $f'$ and $f'_B$.
\footnote{Note that we are working in the so called non-local representation
\cite{Beisert:2006qh}. One can also reintroduce the markers ${\cal
Z}^\pm$ in a simple way.}

Acting with the fermionic generators on both sides we get
constraints on
 $ A,~ B,~ C$, $D,~ E,~ F$, $G,~
H,~ K,~ L$ that  determine them completely up to an overall phase.
We get:
\begin{eqnarray}
A &=& \mathcal{R}_0 \frac{x^+ ( x^+ + x_B)}{x^- ( x^- -
x_B)}\label{r22A}\\ \notag  B &=& \mathcal{R}_0 \frac{  2 x^+ x^- x_B +
(x^+ - x_B) [-2 (x^+)^2 + 2 (x^-)^2 + x^+ x^- ]}{(x^-)^{2}(x^- -
x_B)
}\\\notag
C &=& \mathcal{R}_0  \frac{2 \eta \eta_B }{f} \frac{(x^- + x^+)
(x^- x_B - x^+ x_B - x^- x^+)}{  x_B x^- (x^+)^2
( x^- - x_B )  }\\
D &=& \mathcal{R}_0 \notag \\
E &=&  \mathcal{R}_0 \frac{  2 [ (x^+)^2 - (x^-)^2] [-  x^+ x^- +
x_B ( x^- - x^+ +  x^- (x^+)^2) ] -
 x_B (x^+ x^-)^2 (x_B - x^-)}{(x^- x^+)^{2} x_B ( x^- - x_B )  }\nonumber\\ \notag
F &=& \mathcal{R}_0 \frac{2 f }{\eta \eta_B}\frac{  (x^+ - x^-)
(x^+ + x^-) ( x_B x^+ - x_B x^- + x^+ x^- ) }{ (x^-)^{3} (x^- - x_B) }\\ \notag
G &=& \mathcal{R}_0 \frac{\eta_B}{\eta} \frac{ (x^+ - x^-) (x^+ +
x^-)}{
(x^- - x_B ) x^-}\\ \notag
H &=& \mathcal{R}_0
\frac{(x^+)^2 - x_B x^-   }{x^{-} (x^- - x_B) }\\ \notag
K &=& \mathcal{R}_0 \frac{[x_B x^+ + (x^-)^{2}]}{ x^- (x^- - x_B)}\\ \notag
L &=& \mathcal{R}_0 \frac{\eta }{\eta_B}\frac{ (x^+ +  x^-) x_B }{
x^{-} (x^- - x_B) }
\end{eqnarray}
Notice that the phase $f$ appears explicitly in $C$ and $F$. We can
eliminate $f$ at the cost of introducing markers, ${\cal Z}^\pm$,
 as explained in \cite{Beisert:2006qh}.

The boundary Yang Baxter equation is satisfied by the exactly the  same
argument used by Beisert in \cite{Beisert:2006qh}, as the symmetries
and representations are   the same as in the bulk. As in that
case, there are two intermediate representations for 3 particle
states and each one
contains a state that scatters diagonally.\\
Note also that the boundary scattering in the full theory is given
by taking the product of two such reflection matrices, one for each
$SU(2|2)$ factor.
One could  also derive a  crossing
equation by scattering the identity state \nref{identsta} as we did in the  $SU(1|2)$ case.

Note that  $\frac{A}{D}$ is a prediction for the ratio of amplitudes
of $ Y Y \rightarrow Y Y$ scattering in the $SU(2)$ sector to $ \psi
\psi \rightarrow \psi \psi$ in the $SU(1|1)$ sector. In the
following section we will test the ratio $\frac{A}{B}$ and calculate
the phase factor at weak coupling.

\subsubsection{Boundary bound states}
\label{bdybound}

\begin{figure}
\begin{center}
  \includegraphics[scale=0.6]{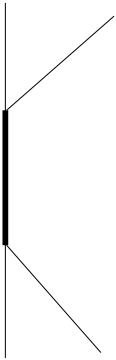}\\
  \caption{Pole at $x^-=x_B$}\label{pole1afig}
  \end{center}
\end{figure}

It is interesting to note that the coefficient $A$ has a pole at
$x^- = x_B$. In the full problem, once we take the product of the two reflection matrices
we expect that the overall phase factor is such that the scattering
in the $SU(2)$ subsector
 continues to have a
single pole at this position. In fact, this will be explicitly
checked at weak coupling in   section \ref{weakzscatt}. Thus, we expect to
have single pole at all values of the coupling. This pole signals
the presence of a bound state, similar to the ones considered in
\cite{Dorey:2006dq}.
Following the same rules as in \cite{Dorey:2007xn}
we see that this pole is a generated by the Landau diagram in figure
\ref{pole1afig} that yields a normalizable wave function.
Figure \ref{pole1afig} represents an actual boundary bound state in
the s-channel. The incoming fundamental magnon binds
to the boundary degree of freedom to form a BPS bound state
corresponding to a double box representation of $SU(2|2)^2$.
 As in the bulk case, we can introduce a
new parameter $x_B^{(2)} \equiv x^+$.   Once we set $x^-=x_B$, we find
that
 \be x_B^{(2)} + { 1
 \over x_{B}^{(2)} } = 2 { i \over g}
 \ee
 The energy of the bound state is given by
 $\epsilon =
 { g \over i} ( x^{(2)}_B - { 1 \over x^{(2)}_B} ) $, as in \nref{enB}.
  We can now consider the boundary scattering of
 another magnon with this new boundary impurity. This can be computed
 by scattering this second magnon, parametrized by $x_2^\pm$,
 off the bound state made out of the    original impurity and
 the first magnon, parametrized by $x^{(2)}_B =x_1^+, ~x^-=x_B$.
 This scattering is described a the product of the scattering amplitudes of the
 second magnon from the first, the reflection matrix, and the scattering of the reflected
 second magnon with the first. This full amplitude
 has  a pole at $x_2^- = x_B^{(2)}  $. Thus we can have a new
 bound state characterized by $x^{(3)}_B\equiv x_2^+$.
  Proceeding in this fashion we obtain
 a structure of bound states very similar to what we had in the bulk
  \cite{Dorey:2006dq,Chen:2006gq}.
 An $n$ particle bound state is given by $x_B = x_1^-$, $x_1^+ = x^-_2$,
 $ x_i^+ = x_{i+1}^-$, $x_B^{(n)} = x_{n-1}^+$. Then using the equations for each of
 the particles one can see that
\be
\label{boundstate} x_B^{(n)} + { 1 \over x_{B}^{(n)} } = n { i \over g} ~,~~~~~~~~
\epsilon_B = { g \over i} ( x^{(n)}_B - { 1 \over x_B^{(2)} } )= \sqrt{ n^2 + 4 g^2 } \ee
These are in the same representation of the extended $SU(2|2)^2$ superalgebra as
the bulk magnons \cite{Chen:2006gp}, except, of course,
that the central charges are given by the line going
from the center of the disk to the rim of the disk.

\section{Results at weak coupling}

In this section we present some results obtained from weak coupling
calculations in the gauge theory. We consider the operators
$\mathcal{O}_Y$ and $\mathcal{O}_Z$ described by expressions
(\ref{oy}) and (\ref{oz}). We   study the large $J$ limit, where the
chain is infinitely long and we focus on the physics near each of
the boundaries. We study  ${\cal N}=4$ super Yang Mills at two
loops, using the results for the dilatation operator obtained in
\cite{Beisert:2003tq} to calculate the reflection matrices in the
$SU(2)$ subsector. Furthermore, we  perform some non trivial checks,
in the $SU(3)$ subsector, of the ratios of the matrix elements of
the exact matrices discussed in the previous section. Finally, in
appendix \ref{apint} we discuss the integrability of the resulting
Hamiltonian.

\subsection{The two loop Hamiltonian at weak coupling in the $SU(2)$ sector}

In order to calculate the reflection matrices we first need to
calculate the appropriate Hamiltonian including the boundary
contributions. This has been calculated at one loop in
\cite{Berenstein:2005vf} and at two loops in \cite{Agarwal:2006gc}.
We review this calculation and discuss an extra term, relative to \cite{Agarwal:2006gc},
 that is present
at two loops. This term, although subtle, is crucial
 to make the spin chain integrable.

Our starting point is the general expression
for the one and two loop dilatation operator
  \cite{Beisert:2003tq} in the SU(2) subsector. This is
\begin{eqnarray}
D = - 2 \frac{g^2}{N} :\textrm{Tr} [Y,Z][\breve{Y},\breve{Z}]: -
2\frac{g^4}{N^2} :\textrm{Tr}
[[Y,Z],\breve{Z}][[\breve{Y},\breve{Z}],Z]:\nonumber\\
- 2\frac{g^4}{N^2}
 :\textrm{Tr}
[[Y,Z],\breve{Y}][[\breve{Y},\breve{Z}],Y]:+ 4\frac{g^4}{N}
:\textrm{Tr} [Y,Z][\breve{Y},\breve{Z}]:\label{opD}
\end{eqnarray}
\noindent where $\breve{X}$ means $\frac{\partial}{\partial X}$.

We can calculate the effective Hamiltonian operating on a $SU(2)$
spin chain from this operator. The bulk part of this Hamiltonian is
 \cite{Agarwal:2006gc,Beisert:2003tq}
\begin{equation}
H_{\textrm{bulk}} = \sum_i (2g^2-8g^4) (I - P_{i,i+1}) + 2 g^4 \sum
(I-P_{i,i+2})
\end{equation}
\noindent where $P_{i,j}$ is the permutation operator between sites
$i$ and $j$.

Let us discuss the boundary terms that need to be added when we
attach our spin chain to a giant graviton. As the interaction has a
range of two sites we only need to worry about the first few sites
of the chain,
assuming a boundary on the left. Let us assume our spin chain starts
as
\begin{equation}
\epsilon X_B^{N-1} | \underbrace{X_0}_{0} \underbrace{X_1}_{1}
\underbrace{X_2}_{2} \ldots
\end{equation}
\noindent where $X_i$ are   fields that can take the values
$Y,Z$. We have been schematic and have omitted indices in this
expression. The $|$ separates the giant graviton from the rest of
the chain.

From the site 1 onwards we have the bulk Hamiltonian. At site 0, the
Hamiltonian acts differently. To leading order in $1/N$,
  the determinant cannot have a field of
the same flavor next to it
\cite{Berenstein:2005vf,Berenstein:2005fa,Berenstein:2003ah}. This
means that $X_0$ is always different from $X_B$. We also have to be
careful about this when we operate with the Hamiltonian. If $X_1$ or
$X_2$ are equal to $X_B$ then the corresponding permutation operator
acting on the site 0 will vanish. With these rules in mind, if we
consider the action of $D$ \nref{opD} on the chain by applying all
derivatives outside the determinant\footnote{This amounts to
truncating $H_{\textrm{bulk}}$ at the end of the chain.}, we find
that $H$ acts on the first three sites as
\begin{equation}
H_{\textrm{naive}} = (2g^2-8g^4) q_1^{X_B} + 2 g^4 q_2^{X_B}
\end{equation}

\noindent where $q_i^{X_B}$ acts as the identity if $X_i = X_B$ and
as zero if it is not. If this was the whole story we would reproduce
the results of \cite{Agarwal:2006gc}. However, we still need to
consider the possibility of the dilatation operator acting on the
determinant and its neighboring sites. It turns out there is only
one term in the dilatation operator \nref{opD} that contributes to
this extra piece. This term is roughly $\frac{g^4}{N^2}\textrm{Tr}
\,\left(\breve{X_B} X_0 X_B \breve{X_B} \breve{X_0} X_B\right)$ with
the first derivative acting on the determinant. Naively, this term
is suppressed by a factor of $N$ as can be seen from (\ref{opD}).
However, since there are $N-1$ letters inside the determinant, there
are $\mathcal{O}(N)$ possible actions of the derivative. All these
subleading terms add up cancelling the $\frac{1}{N}$ suppression.
This extra term is\footnote{This term should also be added to the
expressions in \cite{Okamura:2006zr}.}
\begin{equation}\label{hdet}
H_{\textrm{det}} = 4g^4 q_1^B
\end{equation}
The final form of the two loop boundary Hamiltonian in the $SU(2)$
sector is:
\bea \label{bh}
H &=& H_{\textrm{bulk}} + H_{\textrm{naive}} + H_{\textrm{det}} =
\\ &=&
(2g^2-8g^4) \sum_{i=1}^\infty  (I - P_{i,i+1}) + 2 g^4
\sum_{i=1}^\infty (I-P_{i,i+2}) + (2g^2-4g^4) q_1^{X_B} + 2 g^4
q_2^{X_B} \notag
\eea
Notice that the  chain starts effectively at site 1, as the site 0
is fixed by the boundary\footnote{This situation will change when we
move to the $SU(3)$ subsector}. This Hamiltonian,
with the explicit inclusion of $H_{\textrm{det}}$ \nref{hdet}, is
consistent with integrability.
This is suggested in appendix \ref{apint} by explicitly
constructing the perturbative asymptotic Bethe ansatz solution for
the two magnon problem.

We can now use this result to calculate scattering amplitudes for
different boundaries in the $SU(2)$ subsector.

\subsection{The $SU(1|2)$ reflection matrix off a $\det(Y)$ boundary }

Let us now consider the operators involving an open chain on ending
on the operator $\det(Y)$, corresponding to the $Y=0$ giant graviton
brane. We focus on the large $J$ limit, where we have a large number
of $Z$s producing a long open string, and we focus on one end of the
chain at a time. In that case one can compute the boundary
reflection matrix.
  Let us start considering the operator ${\cal O}_Y$ \nref{yoper} that corresponds to
  the vacuum.
Acting with the Hamiltonian \nref{bh} we find
\begin{equation}
H \mathcal{O}_Y(Z) = 0
\end{equation}
\noindent where we plugged in $X_B=Y$ in the expression (\ref{bh}).
This was expected, since it is a BPS state
the vacuum has zero energy.  We see that we have no degree of freedom, as
the first excitations will be massive. If we place an impurity
moving with momentum $p$ far away from the boundary, all boundary
terms vanish, and we  recover the   bulk expression for the energy
\begin{equation}
H \mathcal{O}_Y(Y_p) = \left(8 g^2 \sin^2 \frac{p}{2} - 32 g^4
\sin^4 \frac{p}{2}\right) \, \mathcal{O}_Y(Y_p)
\end{equation}
\noindent for a one particle state with momentum $p$,
$\mathcal{O}_Y(Y_p)$; see equation \nref{ypdef}. The formula for the
energy is just the expansion to second order in $g^2$ of the
anomalous part of the magnon energy
\begin{equation}
\epsilon - 1 = \sqrt{1+16 g^2 \sin^2 \frac{p}{2}} - 1 \sim 8 g^2
\sin^2 \frac{p}{2} - 32 g^4 \sin^4 \frac{p}{2}
\end{equation}
Let us now compute the reflection matrix.
  We write a wavefunction of the form
\begin{equation}\label{ypdef}
\mathcal{O}_Y(Y_p) = \sum_{x=1}^\infty \Psi(x) \, \mathcal{O}_Y(Y_x)
= \sum_{x=1}^\infty \left(e^{+i p x} + R  e^{-i p x}\right)
\, \mathcal{O}_Y(Y_x)
\end{equation}
\noindent where $\mathcal{O}_Y(Y_x)$ is an operator of the form
given by equation $\ref{oy}$ with the impurity placed at site $x$.
In principle, there can be corrections  of order $g^2$ near $x\sim
0$, as was discussed for the bulk in \cite{Staudacher:2004tk}. This
turns out not to be necessary in our case.
 If we apply the Hamiltonian we
see that this is an eigenstate of the right energy, provided we set
\begin{equation}
\Psi(0) =0 \quad \quad \Psi(-1) + \Psi(1) = 0
\end{equation}
where we have analytically continued the expression for the wavefuntion,
 $\Psi(x) = e^{ip x} + R e^{- ipx}$,
to negative values of $x$.
Remarkably both equations can be satisfied simultaneously without
the inclusion of corrections by setting $R= -1$.
In terms of the reflection matrix for each $SU(1|2)$ factor \nref{rl}, and recalling
the expression for $Y$, \nref{ywexpr}, we see that
\be
 -1 = R = { \cal R}_{0L}^2 e^{2 i p} ~,~~~~~~~~~\to ~~~~{\cal R}_{0L}^2 = - e^{ -2 i p}
 \ee
 up to two loops. We see that the two loop correction vanishes. It would be interesting
 to see at what loop order we get the first deviation from this result.
%
%
%

Finally, we notice that there are no poles associated with boundary
bound states in this matrix. This confirms, at weak coupling, our
assumption that there are no boundary degrees of freedom in this
theory.

\subsubsection{ One loop test  for the $SU(1|2)^2$ reflection
matrix }

In this section we will compare the reflection amplitudes of
 $Y$, $\overline{Y}$ and $W$ ($\overline{W}$ should be the same as
$W$) off a boundary that consists of a $Y = 0$ giant graviton brane.
 These calculations
were performed  at one loop  in \cite{Berenstein:2005vf}, where they
have an expression for the one loop boundary hamiltonian in the
$S0(6)$ sector. In our notation\footnote{Among other things they
define the origin of the chain at site 1 instead of site 0. This
introduces some phases.} the results they obtain for scattering off
a boundary (a $\det(Y)$ boundary) on the left are
\begin{eqnarray}
R_W &=& e^{-i p} =R_{ \overline{W}} \\
R_Y &=& - 1\\
R_{\bar Y} &=& -e^{-2 i p}
\end{eqnarray}

Notice the quotients $\frac{R_W}{R_Y} = - e^{-i p}$ and
$\frac{R_{\bar Y}}{R_Y} = e^{-2 i p}$ are the ones predicted by our
exact matrix (\ref{rl}), recalling the expressions \nref{ywexpr} for
the impurities.
 Also, the
overall factors are the same as the ones calculated at two loops in
this section.

\subsection{The $SU(2|2)$ spectrum and reflection matrix off a $\det(Z)$ boundary }
\label{weakzscatt}

Let us now go through a similar calculation for the $SU(2|2)$
reflection matrix, which corresponds to the case that we have an
open chain ending on a  $\det(Z)$ operator.
 In this case the ground state is non trivial. As we
argued before, the letter placed next to the determinant, $\det(Z)$,
cannot be a $Z$.  This means that, at the very least, one field gets
trapped in between the vacuum described by a chain of $Z$s
 and the D-brane. Our simplest
guess for this operator is $\mathcal{O}_Z(Y,\cdots)$, \nref{oz}, where
the dots represents the other boundary which we are not discussing now.
 Direct computation
shows that this is an eigenstate with energy
\begin{equation}
H \mathcal{O}_Z(Y, \cdots) = (2g^2-2g^4) \, \mathcal{O}_Z(Y,\cdots)
\end{equation}
This energy is the contribution from one boundary. In the case of the full chain,
we have a second impurity at the other end and we have to add the corresponding energy.
This energy agrees precisely with the weak coupling expansion of the
exact formula \nref{enB},
\be
\epsilon_B = \sqrt{1 + 4g^2} \sim 1 + 2 g^2 - 2 g^4
\end{equation}
This computation  tests the boundary term
in the Hamiltonian (\ref{bh}).

Once again, scattering states   have the same energy as in the
bulk, so the total energy is
\begin{equation}\label{en22}
H \mathcal{O}_Z(Y,Y_p,\cdots) = \left[ (2 g^2 - 2 g^4) + (8 g^2 \sin^2
\frac{p}{2} - 32 g^4 \sin^4 \frac{p}{2}) \right] \mathcal{O}_Z(Y,Y_p,\cdots)
\end{equation}
In appendix \ref{apr2} we construct explicitly the wavefunction up
to two loops, check this expression for the energy, and compute the
reflection amplitude to two loops. We find
\begin{eqnarray}\label{r2loop2}
R' = -\frac{1-2 e^{ip}}{1- 2 e^{-ip}} + 2g^2 \frac{e^{- i
p} (e^{ip}-1)^3 (e^{ip} + 1) (1- 4 e^{ip}+ e^{i 2 p})}{(e^{ip}-2)^2}
\end{eqnarray}
This   fixes the overall phase $\mathcal{R}_{0L}$ in
\nref{r22A}
 at
two loops for weak coupling. We would like to write this
expression as a function of $x^\pm, x_B$ such that we can make a
guess that might be correct to a few higher orders as in
\cite{Beisert:2004hm}. Moreover, writing the expression this way
allows for the identification of poles in the reflection matrix in
a straightforward way. Notice that the   coefficient $A$ in the
matrix $\mathcal{R}$ (\ref{r22A}) has the right limit at 1 loop
but disagrees with (\ref{r2loop2}) at two loops.
%
We propose an expression that coincides with \nref{r2loop2} up to two
loops.
\begin{equation}\label{2loopx}
R' = - \frac{\left(x^+ + x_B\right)}{\left(x^- -
x_B\right)}\frac{\left(x^+ + \frac{1}{x_B}\right)}{\left(x^- -
\frac{1}{x_B}\right)}\frac{\left(x^- + x_B\right)}{\left(x^+ -
x_B\right)}\frac{\left(x^- + \frac{1}{x_B}\right)}{\left(x^+ -
\frac{1}{x_B}\right)}
\end{equation}
In checking this it is useful to remember the weak coupling
expansions
\begin{eqnarray}
x_B &=& \frac{i}{2g} \left(1 + \sqrt{1+ 4 g^2}\right) \sim
\frac{i}{g} + i g + \cdots \label{xw1}\\
x^\pm &=& e^{\pm i \frac{p}{2}} \frac{ \left(1 + \sqrt{1+ 16
g^2\sin^2{ p \over 2} }\right) }{ 4 g \sin { p \over 2} } \sim
e^{\pm i \frac{p}{2}} \left(\frac{1}{g\, 2 \sin {p \over 2} } + 2
g\,\sin { p \over 2 } + \cdots \right) \label{xw2}
\end{eqnarray}
%
This expression for $R'$ presents four simple poles. The pole at
$x^- = x_B$ is  responsible for the singularities of the weak
coupling expansion (\ref{r2loop2}). This is the pole that is
already visible at one loop. This pole gives rise to a bound state
in the s-channel and corresponds to the BPS boundary bound states
that we discussed in section \ref{bdybound}. We do not know if all
the other poles of \nref{r2loop2} survive when we add higher order
corrections. It should be possible to perform an analysis similar
to the one in \cite{Dorey:2007xn}, to determine the presence or
absence of the other poles.

We can now also read off the two loop value of ${\cal R}_{0L}$ in \nref{r22A}
\be
\mathcal{R}_{0L} ^2 = { R' \over A^2} =  - \left( \frac{x^{-}}{x^{+ }}\right)^2
\frac{\left(x^-
-x_B\right)}{\left(x^+ - x_B\right)}\frac{\left(x^+ +
\frac{1}{x_B}\right)}{\left(x^+ -
\frac{1}{x_B}\right)}\frac{\left(x^- + x_B\right)}{\left(x^+ +
x_B\right)}\frac{\left(x^- + \frac{1}{x_B}\right)}{\left(x^- -
\frac{1}{x_B}\right)}
\end{equation}

\subsubsection{One  loop test for the $SU(2|2)^2$ reflection
matrix}

We compare  our exact
results  for the reflection matrix, \nref{r22A}, with the weak coupling results,
 as we
did  for the $SU(1|2)^2$ case. Unlike the previous case, this
calculation is not available in the literature. We will need to
compute the scattering process of a $W$ approaching a $Z=0$ brane
with a $Y$ degree of freedom. At one loop the fermions do not play a
role and we can consider the $SU(3)$ sector to be closed. (This can
  be seen from the expression of $C$ in the exact solution,
which  is $\mathcal{O}(g)$ while $A$ and $B$ are of order unity).
Therefore, our process is
\begin{equation}
| Y_B W_p \rangle \rightarrow R'_W | Y_B W_{-p} \rangle + R'_Y |W_B
Y_{-p} \rangle
\end{equation}
The Hamiltonian at one loop for the $SU(3)$ sector can be obtained
by restricting the $SO(6)$ result in \cite{Berenstein:2005vf}. In
our notation this is
\begin{equation}
H = 2 g^2 \left( \sum_{i=0}^\infty (I - P_{i,i+1}) + P_{0,1}
q^Z_1\right)
\end{equation}

This means that when there is $Z$ in the first (1) site it is the
same as in the $SU(2)$ subsector, but the permutation operator does contribute when
$Y$ and $W$ occupy the 0 and 1 site as opposed to the $SU(2)$ case.
The reason for this is obvious: both $Y$ and $W$ can appear next to
the determinant of $Z$s.
We use the following trial eigenstate:
\begin{equation}
\Psi = \sum_{x=1}^\infty \left(e^{i p x} +
R'_W
 e^{- i p
x}\right) |Y_B W_x \rangle +
R'_Y e^{-i p x} | W_B Y_x \rangle
\end{equation}

\noindent where $|X^1_B X^2_x \rangle$ is a state with an $X^1$ at the
boundary (the site labelled by zero)
 and an $X^2$ at position $x$. In the bulk ($x > 1$) the
eigenvalue equation yields the necessary value of the energy for
both $W$ and $Y$ states.
\begin{equation}
E = 2 g^2\left(1 + 2 - e^{i p} - e^{-i p}\right)
\end{equation}
Let us see what happens for the first site
\begin{equation}
E \left(%
\begin{array}{c}
  \psi_W(1) \\
  \psi_Y(1) \\
\end{array}%
\right) = \left(%
\begin{array}{c}
  2 \psi_W(1) - \psi_W(2) - \psi_Y(1) \\
  2 \psi_Y(1) - \psi_Y(2) - \psi_W(1)\\
\end{array}%
\right)
\end{equation}
\noindent where
$\psi_W =e^{i p x} + R'_W e^{- i p x}$ and
$\psi_Y = R'_Y e^{-i p x}$. Using the bulk equations we get
\begin{equation}
\left(%
\begin{array}{c}
  \psi_W(1) \\
  \psi_Y(1) \\
\end{array}%
\right) = \left(%
\begin{array}{c}
   \psi_W(0) - \psi_Y(1)\\
   \psi_Y(0) - \psi_W(1)\\
\end{array}%
\right)
\end{equation}
Plugging the ansatz for the wave function we get
 \begin{eqnarray}
 R'_W &=& \frac{e^{2 i p}-e^{i p} + 1}{e^{i p} - 2}\\
 R'_Y&=& \frac{e^{2 i p}- 1}{e^{i p} - 2}
 \end{eqnarray}
These values satisfy
$|R'_Y|^2 + | R'_W|^2 = 1$ as they should to comply with unitarity.
Now we can compare the quotients
$\frac{ R'_W}{ {R'}}$ and $\frac{ R'_Y}{ {R'}}$with the expected
values from the exact calculations, \nref{r22A}. Here ${R'}$ is the
value encountered in the $SU(2)$ sector at one loop (\ref{r2loop2}).
Namely,
\begin{equation}
R' = \frac{2 e^{i p} -1}{1- 2 e^{- i p}}
\end{equation}
The resulting quotients are:
 \be
 \frac{R'_W}{R'} = \frac{e^{i p} + e^{- i p} - 1}{2 e^{i p} -1}
    ~,~~~~~~~~~~~~~~~~~~
 \frac{R'_Y}{R'} = \frac{e^{i p} - e^{- i p} }{2 e^{i p} -1}
 \ee
 From the exact result \nref{r22A} we have
%
%
 \be
 \frac{R'_W}{ {R'}} = \frac{1}{2} \left(1 +
 \frac{B}{A}\right) ~,~~~~~~~
 \frac{R'_Y}{ {R'}} = \frac{1}{2} \left(1 -
 \frac{B}{A}\right)
 \ee
 Expanding $A,~B$, using the first terms in \nref{xw1}\nref{xw2}, we checked that these
equations are true.
 This is a nontrivial one loop check for the
bosonic subsector of the reflection matrix. A very easy check is that
$R'_Y + R'_W =R'$.

\section{Results at strong coupling}

In this section, we discuss results obtained in the strong coupling
regime from string theory. As long as one is interested in the
leading terms in $g$, it is possible to calculate scattering
amplitudes by calculating time delays in classical sine Gordon
theory \cite{Hofman:2006xt}. We make use of this possibility to
calculate the overall phase of the reflection matrix at strong
coupling for both the $Z=0$ and $Y=0$ giant graviton branes. To be
more precise, at strong coupling there are three regimes, depending
on how we scale the momentum. We can keep the momentum fixed and
then compute as we mentioned above; this is the giant magnon regime.
 We could also scale the momentum as $p \sim 1/g$ and this
corresponds to the near plane wave limit. Finally we can set $p \sim
1/\sqrt{g}$, see \cite{Maldacena:2006rv}. For the case of bulk
scattering it is possible to write a formula which captures the
leading order result both in the plane wave and giant magnon regimes
\cite{afs}. Here we will focus on the giant magnon region. As we
briefly discussed in section \ref{crosssect}, the result in the
plane wave region is trivial. Some results in the near plane wave
region were obtained in \cite{mclias}.

\subsection{Boundary conditions in the sine Gordon theory}

According to the work of Pohlmeyer \cite{Pohlmeyer:1975nb} it is
possible to map the problem of a string propagating on $\mathbb{R}
\times S^2$ into the classical sine Gordon model, see also
\cite{mikhailov}.
 This
connection was used in \cite{Hofman:2006xt} to calculate the strong
coupling limit of the bulk scattering phase of string theory on
$AdS_5 \times S^5$. We will do the same here.

We use string worldsheet coordinates in which $\dot t =1$. Then,
 the
sine Gordon field, $\phi(x,t)$,  is related to the unit vector
$\boldsymbol \eta$ describing the $S^2$ as
\begin{eqnarray}
\cos 2 \phi &=& \dot{\boldsymbol{\eta}}^2 - \boldsymbol{\eta}'^2
\label{sg1}
\end{eqnarray}
\noindent
where
 \be \label{stworel}
\boldsymbol{\eta}^2 =1 ~,~~~~~~~ \dot{\boldsymbol{\eta}}^2 +
\boldsymbol{\eta}'^2 =1 ~,~~~~~~ \dot{\boldsymbol{\eta}} \cdot
\boldsymbol{\eta}' =0
\ee We can consider simple cases leading to different boundary
conditions for the sine Gordon theory.
\begin{enumerate}
    \item Scattering off  a $Z=0$ giant graviton brane \\
    \item Scattering off a $Y=0$ giant graviton brane where we chose the $S^2$ within   brane, e.g.
    the $S^2$ given by $|Z|^2 + (\Phi_1)^2 =1$   \\
    \item Scattering off a $Y=0$ giant graviton brane where we chose the $S^2$ transverse to
     the   brane, e.g.
    the $S^2$ given by $|Z|^2 + (\Phi_3)^2 =1$   \\
\end{enumerate}
Recall that $Z = \Phi^5 + i \Phi^6$, $Y= \Phi^3 + i \Phi^4$.

In the first case the boundary is fixed at the center of the $Z$
plane. This means that the $S^2$ boundary condition is
$\dot{\boldsymbol{\eta}}|_{\textrm{Boundary}} = 0$. Therefore, using
equations (\ref{sg1}) and (\ref{stworel}), we find the Dirichlet
boundary condition    $\phi|_{\textrm{Boundary}} = \frac{\pi}{2}$.
This type of boundary conditions were discussed for the classical
sine Gordon theory in \cite{Saleur:1994yh} and the time delay was
calculated. Note that $\phi = {\pi \over 2} $ corresponds to the
maximum of the sine Gordon potential. This implies that the field
has to move from the maximum to the minimum and this leads to some
energy
that is localized near the boundary. This corresponds to the
boundary degree of freedom, or boundary impurity, that we discussed
above.

The second case represents a string that is entirely contained
inside the D-brane that it is attached to. Therefore, the string end
point (the one ending on the D-brane) can move freely on the $S^2$,
thus $\boldsymbol \eta'=0$ and
 this leads  to another Dirichlet boundary condition for the sine
 Gordon field $\phi|_{\textrm{Boundary}} = 0$.  In this case the field is at the minimum
 of the potential and we have nothing localized at the boundary.

Finally, in the third  case the endpoint of the string, which has to
lie  both on the D-brane and inside the $S^2$,
 has to be on the rim of the disk $|z|=1$, which is the
only region common to both. One can then show that this leads to
%
%
%
%
$\phi'|_{\textrm{boundary}}=0$.

In this fashion, we see how different physical configurations in
$AdS_5 \times S^5$ lead to different boundary problems for the sine
Gordon theory. Interestingly enough, all the boundary conditions
that were discussed belong to the special class that make the
boundary field theory integrable \cite{Ghoshal:1993tm}.
Incidentally, the string theory setup we are studying was shown to
be integrable at large $g$ in \cite{Mann:2006rh}. It would be
interesting to see if other integrable  boundary conditions in the sine
Gordon model  map to other configurations
in the string theory.

We should mention that this description that uses the sine Gordon
theory is only an approximation (valid in the classical limit). It
is not capturing the fact that there are collective coordinates
characterizing the magnon. These arise because the magnon has an
$S^3$ worth of possible orientations inside the $S^5$. (In addition,
we have fermion zero modes \cite{minahanzero}.) As we quantize these
coordinates we get all the BPS bound states with various values of
the angular momentum charge $n$ \cite{Dorey:2006dq,Chen:2006gp}. In
particular, the fundamental impurities, such as the fields $Y,~X$,
etc, have wavefunctions that are spread over this $S^3$. Thus, when
we talked about solutions that were localized within a given $S^2$,
we were making an approximation where we neglected this motion. One
could get a better approximation by considering the solutions in
\cite{sv}, which can be used to describe the classical limit of the
 scattering of BPS bound states
\cite{Dorey:2006dq} with angular momentum $n \sim \mathcal{O}(g)$
from the boundary. In the case of the $Z=0$ brane, where we have a
boundary impurity, we construct the solution as follows. Con
consider a soliton of the bulk theory
 with momentum $p=\pi$ that is at rest at the origin. This is a solution that obeys the
 boundary conditions of the boundary theory.
 Its energy is simply half of the energy of the original soliton. We can
similarly consider the generalizations with angular momentum
discussed in \cite{Chen:2006gp,sv}. In that case both
 the angular momentum and energy are half of what they were in the bulk. However, in the
boundary case, we want to quantize the angular momentum so that it
is an integer after dividing by half. Thus we get a formula for the
energies that has the form \be \epsilon_B = { 1 \over 2} \sqrt{ (2 n
)^2 + 16 g^2 } = \sqrt{ n^2 + 4 g^2 } \ee where $n$ is an integer.
This
 is in agreement with the exact results  \nref{boundstate}.

\subsection{Time delays and scattering phases}

Let us consider first the case where we have a $Y=0$ giant graviton
brane.
 It is convenient to
think about the problem by using a ``method of images'' where the
incoming soliton scatters an antisoliton or a soliton coming from
the other side of the boundary, depending on the boundary
conditions. From our experience with the sine Gordon model and the
bulk calculations in \cite{Hofman:2006xt}, we know the result will
be independent of whether the image state is a soliton or an
antisoliton. Therefore, we don't need to specify this in our
calculations.

When we translate between the sine-Gordon results and the results
computed in the conventions that are more natural at weak coupling
we need to be careful about the fact that these two different
conventions differ in the definition of the spatial coordinate. This
was explained in more detail in
\cite{Hofman:2006xt,Roiban:2006gs,Chen:2006gq}. In fact, we can work
in conventions that coincide with the gauge theory conventions and
notice that the classical boundary scattering amplitude has a simple
relation to the bulk scattering amplitude once we note that the
boundary scattering amplitude can be computed by the ``method of
images''. Let us consider the case where we scatter from a right
boundary\footnote{We can obtain the result for left boundaries by a
parity transformation $\mathcal{R}_L (p) = \mathcal{R}_R (-p)$.}.

For a $Y=0$ brane, we have two solitons, one with momentum $p_1=p$
and another with momentum $p_2 =-p$. The bulk scattering phase is
related to the time delays \be \Delta T_{12} ={ d p_1 \over d
\epsilon_1 }
\partial_{p_1} \delta(p_1,p_2)  ~,~~~~~~~~~~ \Delta T_{21} = { dp_2
\over d \epsilon_2 } \partial_{p_2} \delta(p_1,p_2 ) \ee where
$\delta(p_1,p_2)$ is the bulk scattering phase computed in
\cite{Hofman:2006xt} \be  \label{bulkdelta} \delta(p_1,p_2) = - 4
g\, ( \cos {p_1 \over 2} - \cos{p_2 \over 2} ) \log \left[ { \sin^2{
p_1 - p_2 \over 4} \over  \sin^2{ p_1 + p_2 \over 4} } \right] \ee
where $sign( \sin p_i)>0$. For $p_2 =-p <0$ we should set $p_2 = 2
\pi -p$ in this formula, and this is what we will always mean by
$-p$. In the case that $p=p_1=-p_2$ we find that the two time delays
are equal to each other and to the time delay for scattering from
the boundary $\Delta T_{12} = \Delta T_{21} = \Delta T_B(p)$. Thus
we conclude that the classical (right) boundary scattering phase,
${\cal R}_R = e^{ i \delta_{B,R}}$,
 is the solution to
\bea { d p \over d \epsilon } \partial_p \delta_{B,R}(p) &=& \Delta
T_B(p) = { 1 \over 2 } (\Delta T_{12} +\Delta T_{21} ) = \notag
\\ \label{derivrel}
 &=& { 1 \over 2 } \left. ({ d p_1 \over d \epsilon_1 }
\partial_{p_1} \delta(p_1,p_2)  +
 { dp_2 \over d \epsilon_2 } \partial_{p_2} \delta(p_1,p_2 ) )\right|_{p_1=-p_2=p }
 \eea
 A solution to this equation is then
 \be \label{bdyphase}
 \delta_{B,R}(p) ={ 1 \over 2 }  \delta(p,-p) =
 - 8 g\, \cos { p \over 2} \log \cos { p \over 2 }
 \ee
 where $\delta$ is \nref{bulkdelta}. This describes right-boundary scattering.
Note that we get the same answer regardless of the state of the
impurity, since the matrix structure of the reflection matrix
\nref{rightbdy} is subleading at large $g$. This also means that
this an actual calculation of the overall phase factor
$\mathcal{R}^2_{0R}$ at strong coupling and to leading order.

We can check that this result obeys the classical limit of the
crossing equation \nref{crossfin} \be \delta_{B,R}(p) + \delta_{B,R}(\bar p)
= - \delta(p,-\bar p) + \mathcal{O}(1) \ee where the
$\mathcal{O}(1)$ terms are order one in the $1/g$ expansion. Notice
that in order to get the results for $\bar p$, we should set $p\to
-p$ in \nref{bulkdelta} \nref{bdyphase} and, as we mentioned before,
to get the results for $-p$ we should set $p \to 2 \pi -p$ in
\nref{bulkdelta} \nref{bdyphase}.

This result is valid in the giant magnon regime. We remind the
reader that reflection becomes trivial in the plane wave region, as
magnons become noninteracting. In that case, we get
Dirichlet boundary conditions for the fields $Y, ~\overline{Y}$ and
Neumann for $W,~\overline{W}$. This implies that $\mathcal{R}^2_{0R }= -1$ in
the plane wave regime.

In a similar way we can compute the classical limit of the boundary
scattering for the $Z=0$ brane. In this case we have a boundary
impurity. Using the ``method of images'' we can represent the
boundary impurity  as a third soliton, with momentum $p=\pi$ that is
sitting at the boundary. This type of solutions was obtained
explicitly for the sine Gordon model in \cite{Saleur:1994yh}. In
order to compute right boundary scattering we consider a bulk
configuration with three solitons with $p_1=p$, $p_2 = -p$ and
$p_3=\pi$. Then the time delay is \be \Delta T(p) = \Delta T_{12} +
\Delta T_{13} = { 1 \over 2 } ( \Delta T_{12} + \Delta T_{21} ) +
\Delta T_{13} \ee
 Writing this as in \nref{derivrel} we find the large coupling expression for
 the phase in \nref{r22A},  ${\cal R}^2_{0,R} =e^{i \delta^Z_{B,R}}$,
\be \delta^Z_{B,R}(p) = { 1 \over 2 } \delta(p,-p) + \delta(p,\pi) = -4
g \, \cos {p\over 2 } \, \, \log \left[ \cos^2 {p \over 2 } \, {( 1
- \sin {p\over 2}) \over (1 + \sin {p\over 2 } )} \right] \ee where
\be \label{extrap} \delta(p,\pi ) = - 4 g \, \cos {p\over 2 } \log
\left[ { 1 - \sin {p\over 2} \over 1 + \sin {p\over 2 } } \right]
\ee

The classical limit of the crossing symmetry equation is expected to
be similar and it would still be obeyed since \nref{extrap} is odd
under $p\to -p$ (which is what we should to do to cross $p\to \bar
p$).

\section{Conclusions and discussion}

\subsection{Summary of results}

In this article we considered some D-brane configurations in
$AdS_5 \times S^5$ and considered the worldsheet theory of an open
string ending on the D-brane.
 We focused on the D-branes that correspond to
maximal giant gravitons. In the dual field theory, these D-branes
correspond to determinant operators of the form $\det(Y)$,
$\det(Z)$, where $Y, ~Z$ are two complex combinations of the scalar
fields in ${\cal N}=4$ super Yang Mills.   We considered an open
string attached to this operator with a large value of $J$, where
$J$ is one of the generators of $SO(6)$. In the dual field theory
this corresponds to attaching a long string of $Z$s to the
determinant operator. This can be viewed as a spin chain defined on
an interval.
 We then considered  impurities propagating on this
chain of $Z$s. The symmetries of the problem determine completely
the single impurity reflection matrix up to an overall phase. These
reflection matrices are asymptotic, as in the bulk
\cite{Staudacher:2004tk}. Namely, we need to go far away from the
boundary to measure it. Thus, the strict mathematical definition of
the reflection matrix requires $J=\infty$.

 We considered two cases. First the
case where the determinant operator was $\det(Y)$. In this case the
boundary breaks the bulk symmetry group to an $SU(1|2)^2$ subgroup.
Yet, this symmetry is powerful enough to determine the matrix
structure of the reflection matrix. In fact, in a natural basis, the
reflection matrix is diagonal.

We then considered the case where we have a $\det(Z)$ operator. In
this case an impurity gets trapped between the string of $Z$s
describing the open string ground state and the determinant
operator. This impurity acts as a boundary degree of freedom. This
problem respects the full extended $SU(2|2)^2$ symmetry that we have
on the bulk of a chain of $Z$s, or the bulk of the string in light
cone gauge \cite{fzam}. The boundary impurity transforms in the
fundamental representation of the extended $SU(2|2)^2$ algebra and
has a (complex) central charge with fixed modulus and a phase that
is determined by the momenta of the other particles. This is very
similar to the structure we have in the bulk of the string. The
algebra determines the energy of the boundary impurity. In this
case, the reflection matrix acts on the boundary degree of freedom.
The resulting matrix is rather similar to the one describing the
bulk scattering of two impurities \cite{Beisert:2005tm}. Also, the
bulk particle can form BPS bound states with the boundary degrees of
freedom. Thus, the spectrum of boundary degrees of freedom includes
an index $n$ which characterizes the total number of impurities
forming the bound state.

Both of   reflection matrices obey the boundary Yang Baxter
equation, which is a requisite for integrability. In the first case,
we derived explicitly the form of the crossing equation by
considering the scattering of a particle/hole pair and demanding
that the corresponding reflection amplitude is trivial. This
derivation could be extended to the second case in a straightforward
way.

We then performed computations in the weak coupling regime. Here we
checked the integrability of the system up to two loops. We resolved
the problems raised in \cite{Agarwal:2006gc} by noticing that there
is an extra boundary contribution to the spin chain Hamiltonian. The
results we obtain at two loops are consistent with integrability, in
the sense that the asymptotic Bethe ansatz for two particles works
properly. Nevertheless, we have not proven the full integrability of
the system at two loops. We also computed the undetermined phase
factor in the reflection matrix up to two loops in the weak coupling
expansion. In addition, we checked that the matrix structure
obtained by the symmetry arguments was consistent with the explicit
weak coupling results.

We also computed the strong coupling limit of the reflection phase.
At strong coupling there are two perturbative regimes, the near
plane wave regime and the giant magnon regime, depending on the
momentum of the impurity. We computed the leading order result for
the scattering amplitude in the giant magnon regime. The computation
can be carried out in a simple way by using a ``method of images'',
 where we view the
problem with a boundary in terms of a problem on the full line with
the proper symmetry under reflection\footnote{ This method
is useful for the classical theory but it is
not appropriate for the full quantum theory. }.
This gives the boundary
scattering phase in terms of the bulk scattering phase.

Note that our computations of the matrix structure of the reflection
matrix are valid also for other systems where we have $SU(2|2)$
symmetry. One such system is the plane wave matrix model
\cite{Berenstein:2002jq}, where one can study configurations
analogous to the ones considered
 here,
even though this particular system appears not to be integrable
\cite{beisertni}.

\subsection{Problems for the future}

We would now like to point out to some open directions that seem
worth exploring further.

The most obvious open problem is to find the overall phase factor
by solving the crossing equation, as was done for the bulk in
\cite{Beisertphase}.

Once we know the phase for the two cases, then, one can check that
we get a consistent result by starting with the $\det(Y)$ brane (or
$Y=0$ giant graviton brane) and fill in the vacuum by adding $Y$
impurities until all we have are two $Z$s that get trapped at the
ends. This should correctly reproduce the energy of the ground state
for an open string on a $\det(Z)$ brane (or $Z=0$ giant graviton
brane) containing one impurity at each end. This gives a consistency
check. Alternatively, if we assume it is true, this could give us a
method for computing the  reflection phase for one case once we know
it for the other case.

Once one has found the overall phase, then one can write Bethe
equations that determine the energy of the system. These equations
will describe only the large $J$ limit of the system.  To go to
the limit of small $J$ one will have to use some more clever
methods, which hopefully rely only on the reflection matrix that
we are considering here.
  Some finite $J$ corrections
were computed in \cite{janikconf}, for the closed string case.

It seems possible to study other D-branes in the bulk. For example,
D-branes that are associated to adding flavors to the theory or
D-branes that correspond to adding operators with various
codimensions in the boundary theory. It seems that many of these
cases could be solved by the techniques in this paper, since they
appear to have enough symmetry to completely constrain the
reflection matrix.

Another interesting case to analyze is the situation where we have
local operators on a half BPS Wilson line \cite{Drukker:2006xg}.
When we consider operators with large $J$ we get an open spin
chain. The boundary conditions seem to  preserve a diagonal
$SU(2|2)$ subgroup. This is likely to be enough to fix the
reflection matrix completely.

 \begin{figure}
 \begin{center}
  \includegraphics[scale=.9]{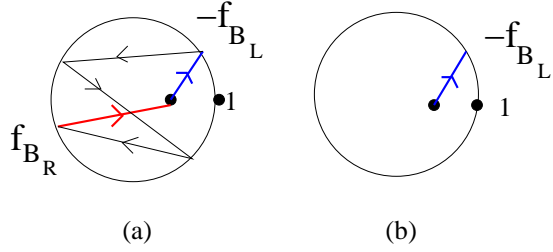}\\
  \caption{(a) Open string configuration with very large $J$ containing three separated magnons which
  ends on a non-maximal giant graviton. In (b) we isolated one of the boundary impurities. Note
  that the length of the boundary impurity line depends on the point along the circle where it
  ends.
  }\label{nonmaximal}
  \end{center}
 \end{figure}

 It seems that one could extend our computations to the case of non-maximal giants, which
 was considered in \cite{Berenstein:2005fa}.
 We again preserve the full extended $SU(2|2)^2$ symmetry, but
 the boundary impurity has a central charge whose absolute value also
 depends on its phase, see figure \ref{nonmaximal}. If we are dealing with a semi-infinite
 chain, then we could compute the matrix structure of the reflection amplitude with the
 methods of this paper. That computation does not rely on integrability. It remains to be
 seen whether the system is integrable or not in this case.

{\bf Acknowledgments }

We would like to thank A. Agarwal and N. Beisert for discussions.
This work was supported in part by DOE grant \#DE-FG02-90ER40542.

\appendix

\section{Appendix: integrability at two loops}\label{apint}

It was pointed out in \cite{Agarwal:2006gc} that the Bethe ansatz
seems to fail at two loops for the problem just studied. We will now
show that the problems raised disappear once we consider the
correct Hamiltonian (\ref{bh}). In particular, the problem was found when
one tried to construct a two particle state using the original
scattering data.

We will consider a wave function of the form $\Psi(x,y) =
\Psi_0(x,y) + g^{2 |x-y|} \Upsilon(x,y)$ where we will only be
concerned with corrections of order $g^2$ to the standard Bethe
ansatz wave function $\Psi_0(x,y)$. This is the asymptotic Bethe
ansatz discussed in \cite{Staudacher:2004tk}. Our state is
\begin{equation}
\mathcal{O}_Y(Y_{p_1} Y_{p_2}) = \sum_{0 < x < y}^\infty \Psi(x,y)
\, \mathcal{O}_Y(Y_x Y_y)
\end{equation}
 The equations we have to
satisfy in the bulk are
\begin{eqnarray}
E \Psi(x,y) &=& (2g^2-8g^4) \left(4 \Psi(x,y) - \Psi(x-1,y) -
\Psi(x+1,y) -\Psi(x,y-1) -\Psi(x,y+1)\right)\nonumber\\
& &+ 2g^4 \left(4 \Psi(x,y) - \Psi(x-2,y) - \right.\nonumber\\
& &\left.\Psi(x+2,y)-\Psi(x,y-2) -\Psi(x,y+2)\right) \quad
\textrm{for}
\quad 2<x<y-2\\
E \Psi(x,x+2) &=& (2g^2-8g^4) \left(4 \Psi(x,x+2) - \Psi(x-1,x+2) -
\Psi(x+1,x+2)\right. \nonumber\\
& &\left. -\Psi(x,x+1) -\Psi(x,x+3)\right) + 2g^4 \left(2 \Psi(x,x+2) - \Psi(x-2,x+2)\right.\nonumber\\
& &\left. -\Psi(x,x+4)\right) \quad \textrm{for}
\quad 2<x \\
E \Psi(x,x+1) &=& (2g^2-8g^4) \left(2 \Psi(x,x+1) - \Psi(x-1,x+1) -\Psi(x,x+2)\right)\nonumber\\
& &+ 2g^4 \left(4 \Psi(x,x+1) - \Psi(x-2,x+1) - \right.\nonumber\\
& & \left. \Psi(x+1,x+2) -\Psi(x-1,x) -\Psi(x,x+3)\right) \quad
\textrm{for} \quad 2<x
\end{eqnarray}
\noindent where $E$ is the sum of the one particle energies. These
equations specify $\Upsilon(x,y)$ completely, as well as the bulk
scattering matrix \cite{Staudacher:2004tk} \cite{Beisert:2005fw}. In
order to obtain information about the reflection matrix we need to
check the eigenvalue equation for sites close to the boundary. If we
pick sites of the form $(2,x)$ our equations are:
\begin{eqnarray}
E \Psi(2,x) &=& (2g^2-8g^4) \left(4 \Psi(2,x) - \Psi(1,x) -
\Psi(3,x) -\Psi(2,x-1) -\Psi(2,x+1)\right)\nonumber\\
& &+ 2g^4 \left(3 \Psi(2,x) - \Psi(4,x) - \right.\nonumber\\
& &\left. \Psi(2,x-2) -\Psi(2,x+2)\right) + 2g^4 \Psi(2,x)\quad
\textrm{for}
\quad 4<x \\
E \Psi(2,4) &=& (2g^2-8g^4) \left(4 \Psi(2,4) - \Psi(1,4) -
\Psi(3,4) -\Psi(2,3) -\Psi(2,5)\right)\nonumber\\
& &+ 2g^4 \left( \Psi(2,4) -\Psi(2,6)\right) + 2g^4 \Psi(2,4)\\
E \Psi(2,3) &=& (2g^2-8g^4) \left(2 \Psi(2,3) - \Psi(1,3) -
\Psi(2,4) \right)\nonumber\\
& &+ 2g^4 \left( 3\Psi(2,3) -\Psi(2,5)- \Psi(3,4) - \Psi(1,2)\right)
+ 2g^4 \Psi(2,3)
\end{eqnarray}
If we use the original equations, these just imply $\Psi(0,x) =
\Psi_0(0,x) = 0$ for $x>2$. These are the analogous equations to
$\Psi(0) =0$ in the single particle case and determine the one
particle reflection matrix to be consistent with the Bethe ansatz.

We still have to consider the sites (1,x). These can't introduce any
more constraints, as our function is already fully determined. The
resulting equations are:
\begin{eqnarray}
E \Psi(1,x) &=& (2g^2-8g^4) \left(3 \Psi(1,x) - \Psi(2,x) -
\Psi(1,x+1) -\Psi(1,x-1)\right)\nonumber\\
& &+ 2g^4 \left(3 \Psi(1,x) - \Psi(3,x) -\Psi(1,x-2) -\Psi(1,x+2)\right) +\nonumber\\
& & (2g^2-4g^4) \Psi(1,x)\quad \textrm{for}
\quad 3<x \\
E \Psi(1,3) &=& (2g^2-8g^4) \left(3 \Psi(1,3) - \Psi(2,3) -\Psi(1,4)
-\Psi(1,2)\right)\nonumber\\
& &+ 2g^4 \left( \Psi(1,3) -\Psi(1,5)\right) + (2g^2-4g^4) \Psi(1,3)\\
E \Psi(1,2) &=& (2g^2-8g^4) \left(\Psi(1,2) - \Psi(1,3)) \right)\nonumber\\
& &+ 2g^4 \left( 2\Psi(1,2) -\Psi(1,4)- \Psi(2,3)\right) +
(2g^2-2g^4) \Psi(1,2)
\end{eqnarray}
Making use of the bulk equations the first of these expressions
yields $\Psi(-1,x) +\Psi(1,x) = \Psi_0(-1,x) +\Psi_0(1,x) = 0$ for
$x>3$. These are the analog of $\Psi(1)+\Psi(-1)=0$ and impose no
further constraints, as our wave function satisfies this identity.
The second equation gives the same result for $x=3$. The last of
these equations is the one that presented a conflict in
\cite{Agarwal:2006gc}. In our case this equation can be written (to
order $g^4$) as
\begin{equation}
2g^4 \left( \Psi_0(1,2) + \Psi_0(-1,2)\right) + (2g^2- 8 g^4)
\Psi_0(0,2) + 2 g^4 \Psi_0(0,1) = 0
\end{equation}
This is satisfied by our Bethe ansatz as $\Psi_0(1,2) +
\Psi_0(-1,2)=0$, $\Psi_0(0,2)=0$ and $\Psi_0(0,1)=0$. This shows
that the two particle problem can be solved by the asymptotic Bethe
ansatz technique, suggesting integrability.

\section{Appendix: computation of the $SU(2|2)$ reflection matrix at two
loops}\label{apr2}

The wave function for a one particle state scattering of the
boundary should satisfy:
\begin{eqnarray}
E \Psi(x) &=& (2 g^2 - 8 g^4) (2 \Psi(x) - \Psi(x+1) - \Psi(x-1))
\nonumber\\
& & + 2 g^4 (2 \Psi(x) - \Psi(x+2) - \Psi(x-2))\nonumber\\
& & + (2g^2-2g^4) \Psi(x) \quad \textrm{for} \quad x>2
\label{bulk22}
\end{eqnarray}
\noindent for the trial wave function $\Psi(x) = \Psi_0(x) + g^2
\delta_{x,1} \Upsilon$. The $g^2$ correction is just an exponential
tail attached to the boundary that accounts for the interactions at
two loops. Further corrections are higher order in $g^2$.
$\Psi_0(x)$ is just the reflecting wave solution $\Psi_0(x) = e^{i p
x} + R' e^{- i p x}$, where $R'$ has, in principle, $g^2$
corrections to the 1 loop result. From this expression we check that
the energy of this state is indeed (\ref{en22}).

The equation that determines $\Upsilon$ comes from the coefficient
of the Schrodinger equation for site 2. Namely
\begin{eqnarray}
E \Psi(2) &=& (2 g^2 - 8 g^4) (2 \Psi(2) - \Psi(3) - \Psi_0(1)) - 2
g^4
\Upsilon \nonumber\\
& & + 2 g^4 (\Psi(2) - \Psi(4)) + (2 g^2- 4 g^4) \Psi(2)
\end{eqnarray}
Using the bulk equation (\ref{bulk22}) we get
\begin{equation}
\Upsilon = \Psi(0) - 2 \Psi(2)
\end{equation}
The equation at site 1 determines the reflection amplitude. This is
\begin{eqnarray}
& &E \Psi_0(1) + 2 g^4 (3-e^{ip} - e^{- i p}) \Upsilon =\nonumber\\
& &(2 g^2 - 8 g^4) (\Psi_0(1)- \Psi(2))  + 2g^4 (\Psi_0(1) - \Psi(3)) +\nonumber\\
& & 2g^4 \Upsilon + 2 g^4 \Psi_0(1)
\end{eqnarray}
\noindent where $2 g^2 (3-e^{ip} - e^{- i p})$ is the one loop
energy extracted from (\ref{en22}). Using the bulk equation we get
\begin{eqnarray}
& &2 g^4 (2-e^{ip} - e^{- i p})\Upsilon = \nonumber\\
& &(10 g^4 - 4 g^2) \Psi_0(1) + (2 g^2 - 8 g^4) \Psi_0(0) + 2 g^4
\Psi_0(-1)
\end{eqnarray}
Plugging in for $\Upsilon$ and the wave function we get
\begin{eqnarray}
& & 2 g^4 (2-e^{ip} - e^{- i p})- 4 g^4 (2 e^{i 2 p} - e^{i 3 p}
- e^{i p}) \nonumber\\
& &- (10 g^4 - 4 g^2) e^{i p} -(2 g^2 - 8 g^4) - 2 g^4 e^{- i p} = \nonumber\\
& &-R' [ 2 g^4 (2-e^{ip} - e^{- i p})- 4 g^4 (2 ^{-i 2 p} - e^{-i p}
- e^{- i 3 p}) \nonumber\\
& &- (10 g^4 - 4 g^2) e^{-i p} -(2 g^2 - 8 g^4) - 2 g^4 e^{ i p}]
\end{eqnarray}
This in turn implies the weak coupling expansion
\begin{eqnarray}
R' = -\frac{1-2 e^{ip}}{1- 2 e^{-ip}} + 2g^2 \frac{e^{- i p}
(e^{ip}-1)^3 (e^{ip} + 1) (1- 4 e^{ip}+ e^{i 2 p})}{(e^{ip}-2)^2}
\end{eqnarray}
This is the result (\ref{r2loop2}).


\begin{thebibliography}{99}

\bibitem{Beisert:2005tm}
  N.~Beisert,
  arXiv:hep-th/0511082.

\bibitem{Janik:2006dc}
  R.~A.~Janik,
  Phys.\ Rev.\  D {\bf 73}, 086006 (2006)
  [arXiv:hep-th/0603038].


\bibitem{Beisertphase}
  N.~Beisert, B.~Eden and M.~Staudacher,
  J.\ Stat.\ Mech.\  {\bf 0701}, P021 (2007)
  [arXiv:hep-th/0610251].
  N.~Beisert, R.~Hernandez and E.~Lopez,
  JHEP {\bf 0611}, 070 (2006)
  [arXiv:hep-th/0609044].

\bibitem{Arutyunov:2006iu}
  G.~Arutyunov and S.~Frolov,
  Phys.\ Lett.\  B {\bf 639}, 378 (2006)
  [arXiv:hep-th/0604043].

\bibitem{Hernandez:2006tk}
  R.~Hernandez and E.~Lopez,
  JHEP {\bf 0607}, 004 (2006)
  [arXiv:hep-th/0603204].


\bibitem{Beisert:2005fw}
  N.~Beisert and M.~Staudacher,
  Nucl.\ Phys.\  B {\bf 727}, 1 (2005)
  [arXiv:hep-th/0504190].

\bibitem{afs}
  G.~Arutyunov, S.~Frolov and M.~Staudacher,
  JHEP {\bf 0410}, 016 (2004)
  [arXiv:hep-th/0406256].

  \bibitem{fzam}
 G.~Arutyunov, S.~Frolov, J.~Plefka and M.~Zamaklar,
  J.\ Phys.\ A  {\bf 40}, 3583 (2007)
  [arXiv:hep-th/0609157].
  G.~Arutyunov, S.~Frolov and M.~Zamaklar,
  JHEP {\bf 0704}, 002 (2007)
  [arXiv:hep-th/0612229].


\bibitem{Berenstein:2002jq}
  D.~Berenstein, J.~M.~Maldacena and H.~Nastase,
  JHEP {\bf 0204}, 013 (2002)
  [arXiv:hep-th/0202021].


\bibitem{Staudacher:2004tk}
  M.~Staudacher,
  JHEP {\bf 0505}, 054 (2005)
  [arXiv:hep-th/0412188].


\bibitem{Santambrogio:2002sb}
  A.~Santambrogio and D.~Zanon,
  Phys.\ Lett.\  B {\bf 545}, 425 (2002)
  [arXiv:hep-th/0206079].

\bibitem{Beisert:2006qh}
  N.~Beisert,
  J.\ Stat.\ Mech.\  {\bf 0701}, P017 (2007)
  [arXiv:nlin.si/0610017].



\bibitem{Ghoshal:1993tm}
  S.~Ghoshal and A.~B.~Zamolodchikov,
  Int.\ J.\ Mod.\ Phys.\  A {\bf 9}, 3841 (1994)
  [Erratum-ibid.\  A {\bf 9}, 4353 (1994)]
  [arXiv:hep-th/9306002].

\bibitem{DeWolfe:2001pq}
  O.~DeWolfe, D.~Z.~Freedman and H.~Ooguri,
  Phys.\ Rev.\  D {\bf 66}, 025009 (2002)
  [arXiv:hep-th/0111135].

\bibitem{McGreevy:2000cw}
  J.~McGreevy, L.~Susskind and N.~Toumbas,
  JHEP {\bf 0006}, 008 (2000)
  [arXiv:hep-th/0003075].

\bibitem{Balasubramanian:2001nh}
  V.~Balasubramanian, M.~Berkooz, A.~Naqvi and M.~J.~Strassler,
  JHEP {\bf 0204}, 034 (2002)
  [arXiv:hep-th/0107119].

\bibitem{Drukker:2006xg}
  N.~Drukker and S.~Kawamoto,
  JHEP {\bf 0607}, 024 (2006)
  [arXiv:hep-th/0604124].

\bibitem{Stefanski:2003qr}
  B.~.~J.~Stefanski,
  JHEP {\bf 0403}, 057 (2004)
  [arXiv:hep-th/0312091].


\bibitem{Chen:2004mu}
  B.~Chen, X.~J.~Wang and Y.~S.~Wu,
  JHEP {\bf 0402}, 029 (2004)
  [arXiv:hep-th/0401016].
  B.~Chen, X.~J.~Wang and Y.~S.~Wu,
  Phys.\ Lett.\  B {\bf 591}, 170 (2004)
  [arXiv:hep-th/0403004].

\bibitem{Susaki:2004tg}
  Y.~Susaki, Y.~Takayama and K.~Yoshida,
  Phys.\ Rev.\  D {\bf 71}, 126006 (2005)
  [arXiv:hep-th/0410139].

\bibitem{mclias}
  T.~McLoughlin and I.~Swanson,
  JHEP {\bf 0608}, 084 (2006)
  [arXiv:hep-th/0605018].

\bibitem{DeWolfe:2004zt}
  O.~DeWolfe and N.~Mann,
  JHEP {\bf 0404}, 035 (2004)
  [arXiv:hep-th/0401041].

\bibitem{Berenstein:2005vf}
  D.~Berenstein and S.~E.~Vazquez,
  JHEP {\bf 0506}, 059 (2005)
  [arXiv:hep-th/0501078].

\bibitem{Berenstein:2005fa}
  D.~Berenstein, D.~H.~Correa and S.~E.~Vazquez,
  Phys.\ Rev.\ Lett.\  {\bf 95}, 191601 (2005)
  [arXiv:hep-th/0502172].


\bibitem{Erler:2005nr}
  T.~G.~Erler and N.~Mann,
  JHEP {\bf 0601}, 131 (2006)
  [arXiv:hep-th/0508064].

\bibitem{Okamura:2005cj}
  K.~Okamura, Y.~Takayama and K.~Yoshida,
  JHEP {\bf 0601}, 112 (2006)
  [arXiv:hep-th/0511139].

\bibitem{Susaki:2005qn}
  Y.~Susaki, Y.~Takayama and K.~Yoshida,
  Phys.\ Lett.\  B {\bf 624}, 115 (2005)
  [arXiv:hep-th/0504209].


\bibitem{Mann:2006rh}
  N.~Mann and S.~E.~Vazquez,
  JHEP {\bf 0704}, 065 (2007)
  [arXiv:hep-th/0612038].

\bibitem{Berenstein:2006qk}
  D.~Berenstein, D.~H.~Correa and S.~E.~Vazquez,
  JHEP {\bf 0609}, 065 (2006)
  [arXiv:hep-th/0604123].


\bibitem{Okamura:2006zr}
  K.~Okamura and K.~Yoshida,
  JHEP {\bf 0609}, 081 (2006)
  [arXiv:hep-th/0604100].

\bibitem{Agarwal:2006gc}
  A.~Agarwal,
  JHEP {\bf 0608}, 027 (2006)
  [arXiv:hep-th/0603067].


\bibitem{Correa:2006yu}
  D.~H.~Correa and G.~A.~Silva,
  JHEP {\bf 0611}, 059 (2006)
  [arXiv:hep-th/0608128].

\bibitem{de Mello Koch:2007uv}
  R.~de Mello Koch, J.~Smolic and M.~Smolic,
  arXiv:hep-th/0701067;
  JHEP {\bf 0706}, 074 (2007)
  [arXiv:hep-th/0701066].

\bibitem{Hofman:2006xt}
  D.~M.~Hofman and J.~M.~Maldacena,
  J.\ Phys.\ A  {\bf 39}, 13095 (2006)
  [arXiv:hep-th/0604135].



\bibitem{Berenstein:2005jq}
  D.~Berenstein, D.~H.~Correa and S.~E.~Vazquez,
  JHEP {\bf 0602}, 048 (2006)
  [arXiv:hep-th/0509015].

\bibitem{Lin:2004nb}
  H.~Lin, O.~Lunin and J.~M.~Maldacena,
  JHEP {\bf 0410}, 025 (2004)
  [arXiv:hep-th/0409174].


\bibitem{Dorey:2006dq}
  N.~Dorey,
  J.\ Phys.\ A  {\bf 39}, 13119 (2006)
  [arXiv:hep-th/0604175].

\bibitem{Chen:2006gp}
  H.~Y.~Chen, N.~Dorey and K.~Okamura,
  JHEP {\bf 0703}, 005 (2007)
  [arXiv:hep-th/0610295].


\bibitem{Corley:2001zk}
  S.~Corley, A.~Jevicki and S.~Ramgoolam,
  Adv.\ Theor.\ Math.\ Phys.\  {\bf 5}, 809 (2002)
  [arXiv:hep-th/0111222].

\bibitem{Balasubramanian:2002sa}
  V.~Balasubramanian, M.~x.~Huang, T.~S.~Levi and A.~Naqvi,
  JHEP {\bf 0208}, 037 (2002)
  [arXiv:hep-th/0204196].



\bibitem{Berenstein:2004kk}
  D.~Berenstein,
  JHEP {\bf 0407}, 018 (2004)
  [arXiv:hep-th/0403110].

\bibitem{Berenstein:2002ke}
  D.~Berenstein, C.~P.~Herzog and I.~R.~Klebanov,
  JHEP {\bf 0206}, 047 (2002)
  [arXiv:hep-th/0202150].



  \bibitem{Berenstein:2003ah}
  D.~Berenstein,
  Nucl.\ Phys.\  B {\bf 675}, 179 (2003)
  [arXiv:hep-th/0306090].

  \bibitem{Das:2000st}
  S.~R.~Das, A.~Jevicki and S.~D.~Mathur,
  Phys.\ Rev.\  D {\bf 63}, 024013 (2001)
  [arXiv:hep-th/0009019].

  \bibitem{Balasubramanian:2004nb}
  V.~Balasubramanian, D.~Berenstein, B.~Feng and M.~x.~Huang,
  JHEP {\bf 0503}, 006 (2005)
  [arXiv:hep-th/0411205].


\bibitem{Dorey:2007xn}
  N.~Dorey, D.~M.~Hofman and J.~Maldacena,
  Phys.\ Rev.\  D {\bf 76}, 025011 (2007)
  [arXiv:hep-th/0703104].


\bibitem{Beisert:2003tq}
  N.~Beisert, C.~Kristjansen and M.~Staudacher,
  Nucl.\ Phys.\  B {\bf 664}, 131 (2003)
  [arXiv:hep-th/0303060].

\bibitem{Beisert:2004hm}
  N.~Beisert, V.~Dippel and M.~Staudacher,
  JHEP {\bf 0407}, 075 (2004)
  [arXiv:hep-th/0405001].


\bibitem{Maldacena:2006rv}
  J.~Maldacena and I.~Swanson,
  Phys.\ Rev.\  D {\bf 76}, 026002 (2007)
  [arXiv:hep-th/0612079].


\bibitem{Pohlmeyer:1975nb}
  K.~Pohlmeyer,
  Commun.\ Math.\ Phys.\  {\bf 46}, 207 (1976).


\bibitem{mikhailov}
  A.~Mikhailov,
  arXiv:hep-th/0511069;
  A.~Mikhailov,
  arXiv:hep-th/0507261;
  A.~Mikhailov,
  J.\ Geom.\ Phys.\  {\bf 56}, 2429 (2006)
  [arXiv:hep-th/0504035].



\bibitem{Saleur:1994yh}
  H.~Saleur, S.~Skorik and N.~P.~Warner,
  Nucl.\ Phys.\  B {\bf 441}, 421 (1995)
  [arXiv:hep-th/9408004].

\bibitem{minahanzero}
  J.~A.~Minahan,
  JHEP {\bf 0702}, 048 (2007)
  [arXiv:hep-th/0701005].

\bibitem{sv}
  M.~Spradlin and A.~Volovich,
  JHEP {\bf 0610}, 012 (2006)
  [arXiv:hep-th/0607009];
C.~Kalousios, M.~Spradlin and A.~Volovich,
  JHEP {\bf 0703}, 020 (2007)
  [arXiv:hep-th/0611033].



\bibitem{Roiban:2006gs}
  R.~Roiban,
  JHEP {\bf 0704}, 048 (2007)
  [arXiv:hep-th/0608049].

\bibitem{Chen:2006gq}
  H.~Y.~Chen, N.~Dorey and K.~Okamura,
  JHEP {\bf 0611}, 035 (2006)
  [arXiv:hep-th/0608047].


\bibitem{beisertni}
  N.~Beisert and T.~Klose,
  J.\ Stat.\ Mech.\  {\bf 0607}, P006 (2006)
  [arXiv:hep-th/0510124].




\bibitem{janikconf}
R. Janik, talk at the 12th Itzykson meeting, Paris 2007,
http://www-spht.cea.fr/Meetings/Rencitz2007/agenda.php

















































\end{thebibliography}
\end{document}